\documentclass[amssymb,amsmath,preprint,nofootinbib,superscriptaddress,aps]{revtex4-1}
\usepackage[colorlinks]{hyperref}
\usepackage{graphicx}
\usepackage{multirow}

\usepackage{slashed}
\usepackage{gensymb}
\usepackage{orcidlink}
\usepackage{subfigure}
\usepackage{float}
\usepackage{microtype}  

\begin{document}
\title{Constraining axial non-standard neutrino interactions with MINOS and MINOS+}

\author{S. Abbaslu\, \orcidlink{0000-0003-3567-717X}}
\email{s-abbaslu@ipm.ir}
\affiliation{School of Physics, Institute for Research in Fundamental Sciences (IPM), Tehran, Iran}

\author{Y. Farzan\, \orcidlink{0000-0003-0246-5349}}
\email{yasaman@theory.ipm.ac.ir}
\affiliation{School of Physics, Institute for Research in Fundamental Sciences (IPM), Tehran, Iran}

\begin{abstract}
   We show that the neutral current data of the MINOS and MINOS+ experiments can provide information on the axial neutral current non-standard interactions of neutrinos with the $u$ and $d$ quarks; {\it i.e.,} on $\epsilon_{\alpha \beta}^{Aq}$. We derive bounds on the $ee$, $e\tau$ and $\tau \tau$ components of these couplings and show that the MINOS(+) bounds on $\epsilon^{Aq}_{e\tau}$ and $\epsilon^{Aq}_{\tau\tau}$ are currently the world leading ones. The bound on the isospin singlet case, $\epsilon^{Au}_{\tau\tau}=\epsilon^{Ad}_{\tau\tau}$  is of particular interest because while this isospin singlet NSI is theoretically motivated, it was practically unconstrained before these results.
\end{abstract}

\maketitle

\section{Introduction}\label{sec:intro}
In 1973, the discovery of neutral current interactions of neutrinos with electrons in the Gargamelle detector at CERN played a crucial role in establishing the $SU_L(2) \times U(1)$ gauge symmetry as the underlying theory for the electroweak interactions, leading to the 1979  Nobel prize being awarded to the founders of this theory, S. Glashow, S. Weinberg and A. Salam. If  there are new particles coupled to both neutrinos and matter fields, they can lead to Non-Standard Interactions (NSI) of the neutrinos with the matter fields. In the context of neutrino physics, NSI is usually referred to  effective four-Fermi interactions between neutrinos and the matter fields ({\it i.e.,} electrons and light quarks).  In recent years, the developments of various precise neutrino experiments have enhanced the interest in NSI both from observational and model building points of view. Like the standard weak interactions, NSI can be divided into two classes: the charged current interactions and the Neutral Current (NC) interactions. Neutral current NSI with large effective coupling comparable to the Fermi constant can originate from light new neutral mediators  (lighter than $m_W$) coupled to both neutrinos and matter fields \cite{Babu:2019mfe,Farzan:2016wym,Farzan:2015hkd,Farzan:2015doa,Bischer:2018zbd,Forero:2016ghr,Farzan:2019xor}. In other words, searching for the potential effects of NC NSI can shed light on the existence of  new light particles yet to be discovered. Because of such an intriguing possibility, the NC NSI is the focus of this paper.

The neutral current  non-standard interaction contribution to the effective Lagrangian can be defined as  the following four-Fermi interaction between neutrinos and the  matter fields, $f \in \{ e, u, d\}$:
\begin{equation}
	V_{NSI}=\frac{G_F}{\sqrt{2}}[\bar{\nu}_\alpha \gamma^\mu (1-\gamma^5)\nu_\beta][\bar{f}\gamma_\mu(\epsilon_{\alpha\beta}^{Vf}+\epsilon_{\alpha\beta}^{Af}\gamma^5)f]  \  ,
	\label{eff}\end{equation} where
$\epsilon_{\alpha\beta}^{Vf}$ and $\epsilon_{\alpha\beta}^{Af}$ are vector and axial non-standard interaction couplings. Since the Fermi constant is factored out in Eq. (\ref{eff}),
$\epsilon_{\alpha\beta}^{Vf}$ and $\epsilon_{\alpha\beta}^{Af}$ are  both dimensionless. In the limit $\epsilon_{\alpha\beta}^{Vf}, \epsilon_{\alpha\beta}^{Af}\to 0$, we recover the standard model of particles.

The vector NSI, $\epsilon^{Vf}$,   can affect the neutrino propagation in matter ({\it i.e.,} forward scattering of the neutrinos off the electrons and nucleons composing the medium). Moreover, 
$\epsilon_{\alpha\beta}^{Vf}$ and $\epsilon_{\alpha\beta}^{Af}$  can affect the Coherent Elastic $\nu$ Nucleus Scattering (CE$\nu$NS). Invoking these features, strong bounds on $\epsilon^{Vf}$ are extracted from various neutrino oscillation and CE$\nu$NS experiments.
The axial NSI cannot affect the neutrino propagation in matter or CE$\nu$NS but they can of course appear in the  cross section of the neutral current scattering of neutrinos off a target. 
We shall review the existing bounds on $\epsilon^{Aq}$ ($q \in \{ u,d\})$ in the next section.
For the isospin invariant case, $\epsilon^{Au}=\epsilon^{Ad}$, the $ee$, $e \tau$ and $\tau \tau$ components can be as large as 1. That is the NSI of $\nu_e$ and $\nu_\tau$ can be as strong as the weak interaction. Moreover even for $\epsilon^{Au}=-\epsilon^{Ad}$, there is a non-trivial solution with $\epsilon_{\tau \tau}^{Ad}=-\epsilon_{\tau \tau}^{Au}=1$. The NSI of the $s$ quark is also practically unconstrained.

In Ref. \cite{Abbaslu:2024hep},  a model with rich phenomenology is presented that can lead to $\epsilon_{\tau\tau}^{Aq} \sim 1$. Producing a $\nu_\tau$ or $\bar{\nu}_\tau$ beam to test $\epsilon_{\tau \tau}^{Aq}$ or  $\epsilon_{\tau e}^{Aq}$
is  challenging; however, the neutrino beam at the far detector of a long baseline experiment such as MINOS or DUNE  has a significant $\nu_\tau$ or $\bar{\nu}_\tau$ component. Invoking this feature, Ref. \cite{Abbaslu:2023vqk} shows that the far detector of DUNE can constrain  $\epsilon_{\tau \tau}^{Aq}$  and  $\epsilon_{\tau e}^{Aq}$ down to $\sim 0.1$. In this paper, we use the data of the neutral current events collected by MINOS and MINOS+ for the $\nu_\mu$ mode to search for the impact of  $\epsilon_{\alpha \beta}^{Aq}$,  with $q \in \{ u,d\}$ \footnote{In  Refs. \cite{Blennow:2007pu,Friedland:2006pi,MINOS:2013hmj, Kitazawa:2006iq,Kopp:2010qt,Isvan:2012zia, Coelho:2012bp},  constraining the vector NSI by the MINOS data has already been extensively  scrutinized. However, the potential of the neutral current data for constraining the axial NSI has not been explored up to now.}. We show that the bounds on  $\epsilon_{\tau \tau}^{Aq}$  and  $\epsilon_{\tau e}^{Aq}$  can be significantly improved.
We also discuss the possibility to constrain $\epsilon^{As}_{ \alpha
\beta}$ and $\epsilon^{Vs}_{ \alpha\beta}$ with the NC data from  MINOS(+).

This paper is organized as follows. The present bounds in the literature on the axial NSI are enumerated in sect. \ref{sec:Bounds}. A short description of MINOS and MINOS+ as well as the definition of $\chi^2$ statistics are provided in sect. \ref{sec:MINOS}. The effects of the axial NSI on the NC events are discussed and formulated in sect. \ref{sec:NC}. Our bounds from MINOS and MINOS+ are presented in sect. \ref{sec:OURS}. Results are summarized in sect. \ref{sec:Sum} and then a brief road-map for further studies in this direction is depicted in this section.

\section{Previous Bounds on NSI \label{sec:Bounds}}
The axial NSI $\epsilon_{\mu \alpha}^{Aq}$ with $q \in \{u,d\}$ are severely constrained by  the NuTeV experiment: $|\epsilon_{\mu \mu}^{Aq}|<0.01$ and  $|\epsilon_{\tau \mu}^{Aq}|<0.1$ \cite{NuTeV:2001whx}. On the $\mu e$ component, there is even a more stringent bound from a loop-induced contribution to $\mu+{\rm Ti} \to e +{\rm Ti}$:  $|\epsilon_{e \mu}^{Aq}|<7.7\times 10^{-4}$ \cite{Davidson:2003ha}.  However, the $ee$, $e\tau$ and $\tau \tau$ components are less constrained. The CHARM neutrino scattering experiment constrains the absolute values of the $ee$ and $e\tau$ components to be smaller than 1 \cite{Davidson:2003ha}. Moreover, the measurement of the neutral current events induced by the solar neutrinos at SNO can constrain the difference between  the axial couplings to the $u$ and $d$ quarks  \cite{Coloma:2023ixt}
$$ -0.19<\epsilon_{ee}^{Au}-\epsilon_{ee}^{Ad}<0.13 \ \ {\rm and } \ \   -0.13<\epsilon_{e\tau}^{Au}-\epsilon_{e\tau}^{Ad}<0.1 \ .$$
Ref. \cite{Coloma:2023ixt} also
finds non-trivial  disconnected solutions for $\epsilon^{Au}_{e\tau}-\epsilon^{Ad}_{e\tau}$
which  are already at tension with the CHARM bounds (see also  Ref.~ \cite{Gehrlein:2024vwz}).
Along with a solution around zero, Ref \cite{Coloma:2023ixt} has found a non-trivial solution as
\begin{equation}-2.1<\epsilon_{\tau \tau}^{Au}-\epsilon_{\tau \tau}^{Ad}<-1.8 \ \  {\rm or} \ \  -0.2 <\epsilon_{\tau \tau}^{Au}-\epsilon_{\tau \tau}^{Ad}<0.15 \ .\label{nontrivial}
\end{equation}
 We will show that the MINOS(+) data can rule out these non-trivial disconnected solutions with high confidence level.

As discussed before, from the neutrino oscillation pattern in matter and CE$\nu$NS, there are strong bounds on the vector NSI of the valence quarks of nucleon; {\it i.e.,} on $\epsilon^{Vu}_{\alpha \beta}$
and $\epsilon^{Vd}_{\alpha \beta}$. However, the $s$ quark is not a valence quark of the proton and neutron. Thus, the neutrino oscillation experiments or CE$\nu$NS cannot constrain $\epsilon^{As}_{\alpha \beta}$
and $\epsilon^{Vs}_{\alpha \beta}$.   These couplings can be as large as $O(1)$. The model in Ref. \cite{Abbaslu:2024hep} predicts $\epsilon_{\tau \tau}^{Vs} \sim 1$ and $\epsilon^{As}_{\alpha \beta}=0$.  In fact, their impacts on quasi-elastic and resonance scattering are negligible, too \cite{Abbaslu:2024jzo} but they may leave a discernible impact on DIS \cite{Abbaslu:2023vqk}.
 As shown in  \cite{Abbaslu:2023vqk}, $\epsilon^{As}_{\tau \alpha}$ 
 and $\epsilon^{As}_{\mu \alpha}$ can be significantly improved by a DUNE-like experiment. In this paper, we shall study  the possibility of constraining $\epsilon^{As}_{\alpha \beta}$ and  $\epsilon^{Vs}_{\alpha \beta}$ 
 by the existing MINOS and MINOS+ data.
 
\section{MINOS and MINOS+\label{sec:MINOS}}
The MINOS experiment was a long baseline neutrino experiment which invoked the NUMI beam from FermiLab \cite{MINOS:2017cae}, collecting data from  2005 to 2012. 
MINOS had two detectors made of steel planes interleaved with scintillator strips. The near and far detectors were respectively  located at distances of 1.04 km and 735 km from 
the source. The far detector was installed in the Soudan underground mine in Minnesota. The neutrino beam for MINOS was peaked at 3~GeV. In 2013,  MINOS was followed by MINOS+ with the peak energy at 7~GeV.
The main focus of the MINOS and MINOS+ collaboration was to study the Charged Current (CC) scattering of neutrinos off the nuclei to extract information on the oscillation parameters. However, they have also studied the Neutral Current (NC) scattering data.  Nonzero $\epsilon^{Aq}$ can  only affect NC events so we shall focus on the NC events in this paper.

\begin{figure*}[t!]
    \centering
       \subfigure[]{ \includegraphics[width=0.48\textwidth, keepaspectratio]{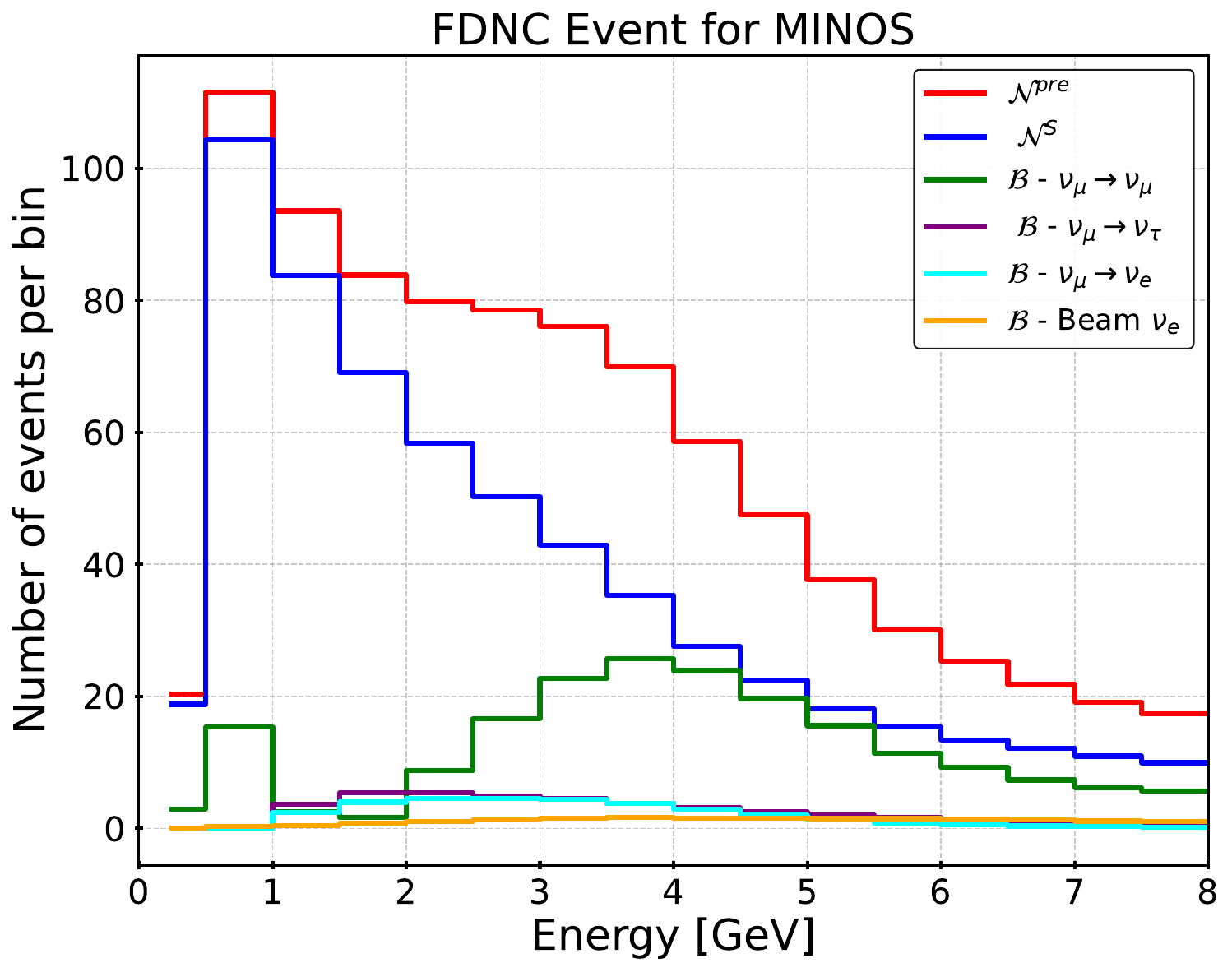}\label{FDNC_components-minos-a}}
        \subfigure[]{\includegraphics[width=0.48\textwidth ]{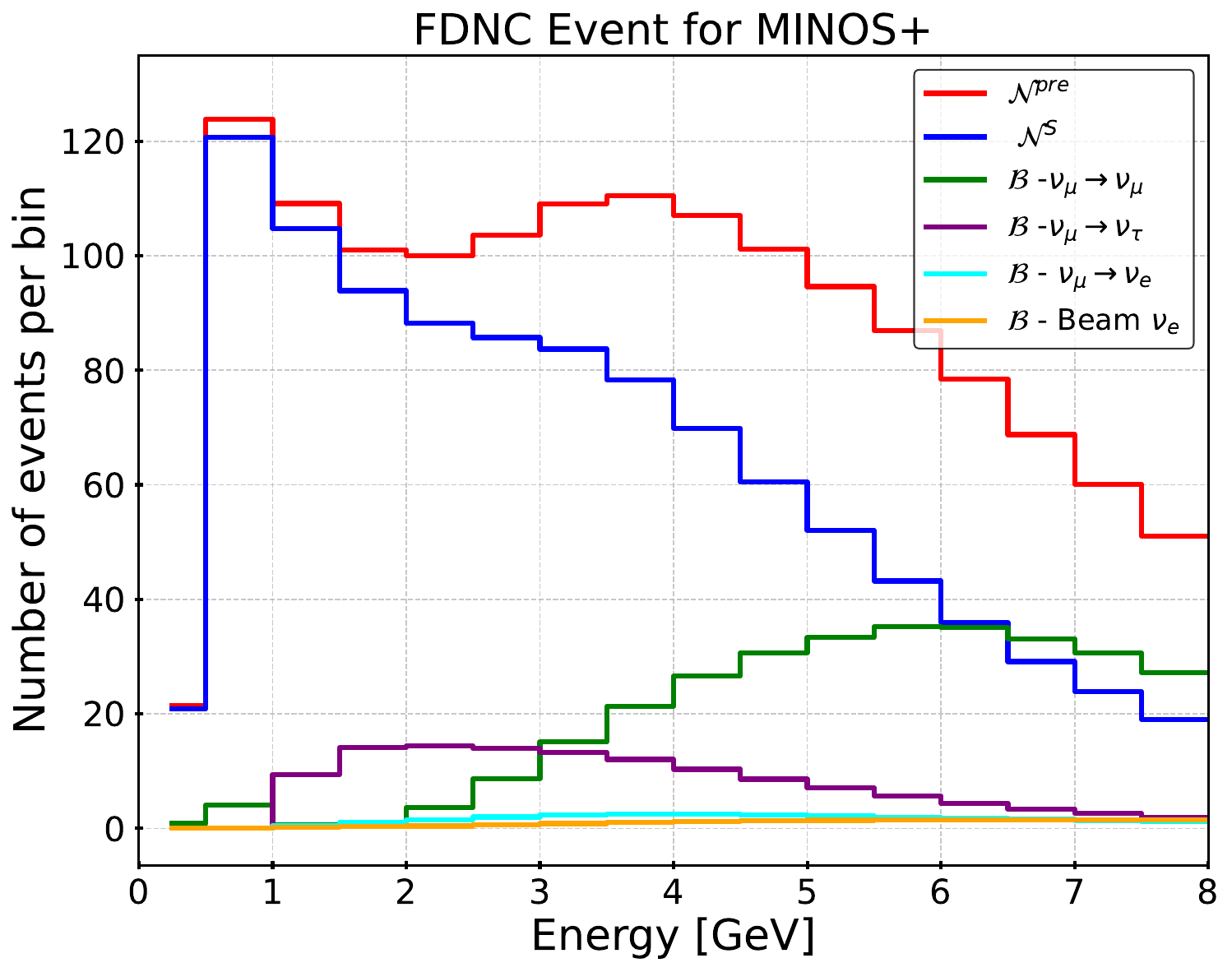}\label{FDNC_components-minosplus-b}}

       \subfigure[]{ \includegraphics[width=0.48\textwidth ]{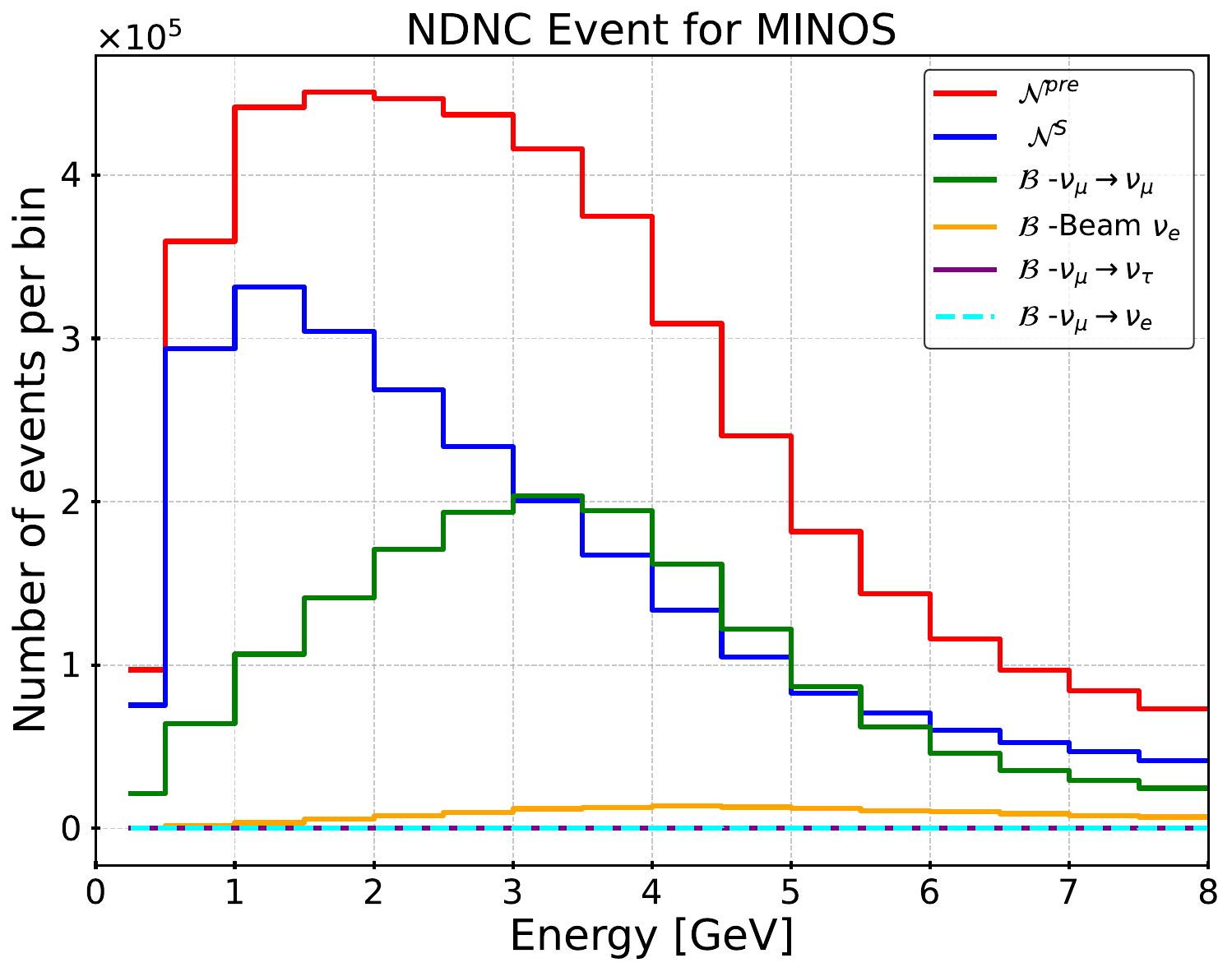}\label{NDNC_components-minos-c}}
    \subfigure[]{\includegraphics[width=0.48\textwidth ]{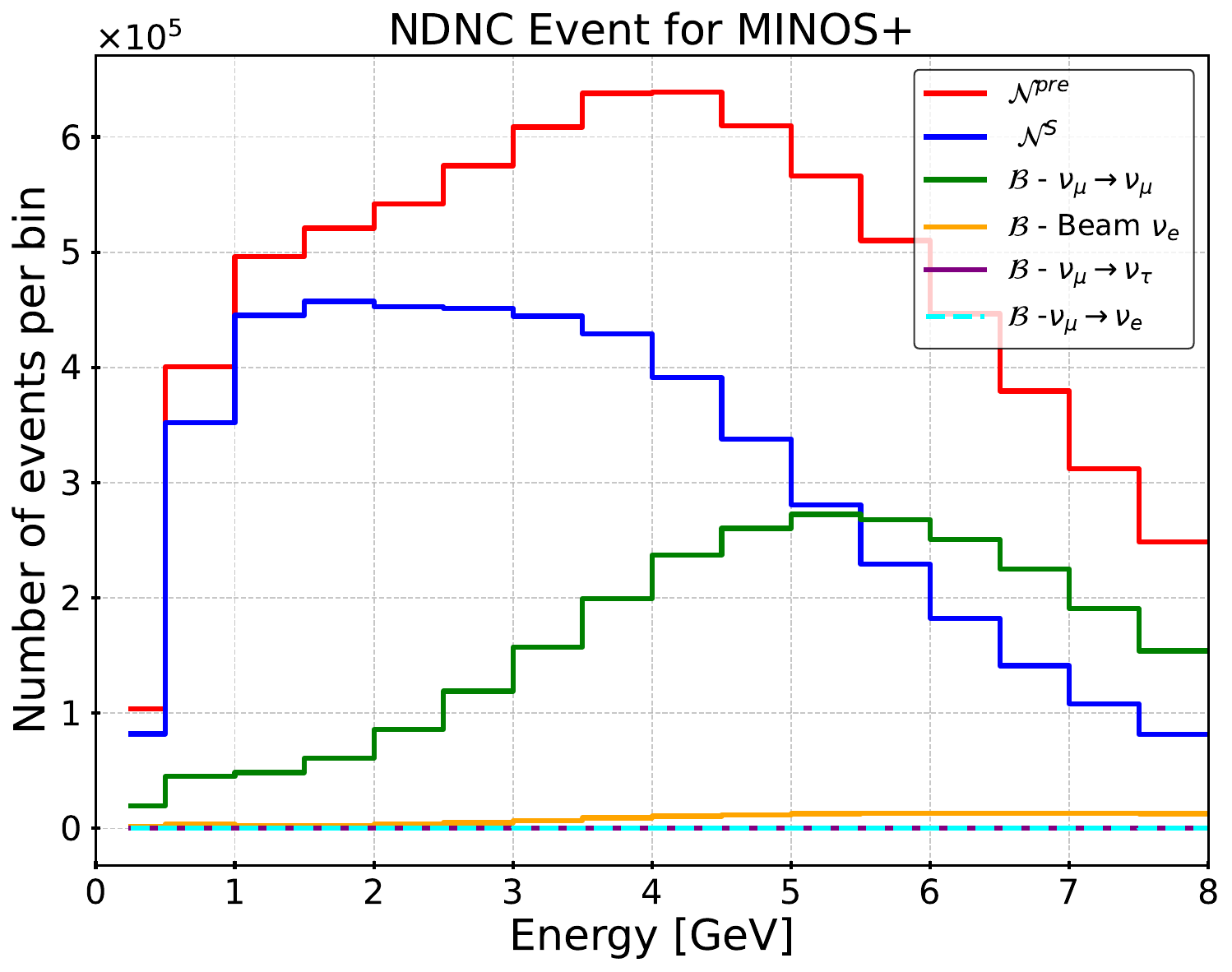}\label{NDNC_components-minosplus-d}}
     \vspace{0.em}
 \caption{Number of predicted NC events and its backgrounds per bin. The red histograms show the total number of events per bin ({\it i.e.,} signal plus all backgrounds). The blue line shows the prediction for the NC signal. The green, violet and cyan histograms respectively show the backgrounds from the misidentification of CC interactions of $\nu_\mu$, $\nu_\tau$ and $\nu_e$ from the flavor oscillation of the beam. The orange lines show the misidentified CC interaction of the intrinsic $\nu_e$ in the beam from the source. The upper (lower) panels are the prediction for the far (near) detector. The left (right) panels are the predictions for MINOS (MINOS+). The width of the shown bins is 0.5 GeV. 
 }\label{fig:ND-FD-Minos-p-NC}
\end{figure*}

MINOS(+) had both neutrino and antineutrino runs but the publicly available NC data only consists of the neutrino mode with the total  16.36$\times 10^{20}$ Protons On Target (POT). We shall therefore only use the neutrino mode data; however, we should emphasize that the NC data from the antineutrino mode could lead to improved bounds on $\epsilon^{Aq}$.    
In the NC scattering, a significant fraction of the neutrino energy is carried away by the final neutrino. The initial neutrino energy is related to the measured (deposited) energy through the so-called migration matrices. The information related to these matrices is incorporated in the histograms that are publicly available in the supplementary material of \cite{MINOS:2017cae}.  The histograms include 27 energy bins for each of the near and far detectors. The bin energies span from 0 to 40 GeV. 
 Fig \ref{fig:ND-FD-Minos-p-NC} shows the NC events along with the various contributions to the background predicted for both the near and far detectors of MINOS and MINOS+. We clip the figures at  8 GeV above which the data is scarce. In our analysis, we however invoke all the 54 bins up to 40 GeV represented in \cite{MINOS:2017cae}. The widths of shown bins are 0.5 GeV. The red histogram depicts the predicted total number of events per bins  which is the sum of the predicted signal shown by the blue histogram and the contributions from all the backgrounds.
 Within the SM, NC is flavor universal so the signal prediction (shown by blue) does not depend on the neutrino oscillation parameters. However, since NSI is in general non-universal in the flavor space, the prediction for the signal in the presence of nonzero $\epsilon^{Aq}$ will depend on the neutrino oscillation parameters.

The background for the signal is mainly composed of misidentified CC  interaction. The neutrino beam  reaches the near detector almost unoscillated so it dominantly consists of  $\nu_\mu$ plus a small $\nu_e$  ($O(10^{-3}$)) sub-component mainly from the three-body Kaon decay at the source. The contributions from these two are shown respectively with green and orange histograms.  Thanks to the distinct track of the muon, the probability of  misidentifying a muon neutrino CC event is low so the background from $\nu_\mu$ CC  is  suppressed. The background from each component is presented in \cite{MINOS:2017cae} which we shall use in our analysis.
At the far detector, we can also have a background from the misidentified CC interaction of $\nu_\tau$ and $\nu_e$ that appear  in the beam due to oscillation.  The background from $\nu_e$ and $\nu_\tau$ are shown with violet and cyan histograms, respectively. Naturally, the background in the far detector depends on the neutrino oscillation parameters but  since the background  originates from  the CC, it does not depend on the NC NSI couplings that we study in this paper. The background from each component is given in the supplementary material of \cite{MINOS:2017cae}. 

Various sources of systematic uncertainties have been taken into account in \cite{MINOS:2017cae}. The most important  among them are the uncertainties in the computation of the  following:  1) hadron production at the source which determines the energy spectrum and the flavor composition of the neutrino beam at the source; 2) scattering cross section; 3) energy scale; 4) beam optics; 5) detector acceptance; 6) effective normalization. Most of these uncertainties ({\it e.g.,} those of the cross section or normalization) are correlated between the far and near detectors. Ref. \cite{MINOS:2017cae} treats all the uncertainties, including both statistical and systematic,  with a covariance matrix that is available in its supplementary material.  Following Ref. \cite{MINOS:2017cae}, we define
\begin{equation}  \label{ChiNC}
	\chi^2_{NC}=\sum_{i=1}^N \sum_{j=1}^N (\mathcal{N}_i^{obs}-\mathcal{N}_i^{pre}(\epsilon))(V^{-1})_{ij}(\mathcal{N}_j^{obs}-\mathcal{N}_j^{pre}(\epsilon)) \ ,
	\end{equation}
where $N$ is the total number of  Near Detector (ND) and Far Detector (FD) bins.  $(V^{-1})_{ij}$ is the inverse of the covariance matrix, which incorporates  both statistical and systematic uncertainties. $\mathcal{N}_i^{obs}$ is the observed NC events in the $i$th bin and $\mathcal{N}_i^{pre}(\epsilon)$ is the predicted value for the  $i$th bin with NSI coupling $\epsilon$. 
$\mathcal{N}_i^{pre}(\epsilon)$ includes both the signal, $\mathcal{N}_i^{S}(\epsilon)$, and the background, $\mathcal{B}_i$:
$$\mathcal{N}_i^{pre}(\epsilon)=\mathcal{N}_i^{S}(\epsilon)+\mathcal{B}_i \ . $$
In the next section, we discuss how to compute $\mathcal{N}_i^{S}(\epsilon)$. As explained before, both $ \mathcal{N}_i^{S}(\epsilon)$ and  $\mathcal{B}_i$ depend  on the neutrino mixing parameters which suffer from extra uncertainties that are not accounted for in the covariance matrix. To take into account the uncertainties of the neutrino mass and mixing parameters, we evaluate $\chi^2_{NC}$ marginalizing over the neutrino parameters with a Gaussian prior.

To constrain $\epsilon$,  we compute $\chi_{NC}^2$ as a function of  nonzero NSI couplings  and find its minimum $\chi^2_{NC}|_{min}$. We then define
\begin{equation} \label{Delta-chi}
	\Delta \chi^2(\epsilon)\equiv  \chi^2_{NC}(\epsilon)
-\chi^2_{NC}|_{min}
\end{equation}
where $\chi^2_{NC}|_{min}$ is the minimum over varying the nonzero $\epsilon$ components as well as the neutrino mixing parameters within the present uncertainties.

\begin{table*}[htb]
\begin{tabular*}{\textwidth}{@{\extracolsep{\fill}}lrrrrrrl@{}}
\hline
$\theta_{12}/\degree$ & $\theta_{23}/\degree$ &  $\theta_{13}/\degree$ & $\delta/\degree$ & $\Delta m_{12}^2/10^{-5} {\rm eV^2}$ & $\Delta m_{31}^2/10^{-3} {\rm eV^2}$ \\
\hline
$33.68^{+0.73}_{-0.70}$ & $48.5^{+0.8}_{-0.9}$ & $8.52^{+0.11}_{-0.11}$ & $177^{+19}_{-20}$ & $7.49^{+0.19}_{-0.19}$ & $2.534^{+0.025}_{-0.023}$ \\
\end{tabular*}
\caption{Three flavor neutrino oscillation parameters for Normal mass Ordering (NO) taken from NuFIT 6.0 global fit \cite{Esteban:2024eli}. These values are obtained without including atmospheric neutrino data from the Super-Kamiokande collaboration (SK-atm).}\label{tab:mixing-parameters}
\end{table*}

\section{Neutral current events at MINOS and MINOS+ with NSI \label{sec:NC}}
In Ref. \cite{MINOS:2017cae}, the MINOS collaboration has used the MINOS and MINOS+ data with a total exposure of $16.36 \times 10^{20}$ Protons-On-Target (POT) to probe the $3+1$ model. In this analysis, a two-detector fit method is invoked. That is instead of using the ratio of the events at the far over those at the near detectors, the reconstructed energy spectra of both near and far detectors are fitted to the sterile neutrino model.  We adopt a similar approach using the supplementary material of \cite{MINOS:2017cae} which includes the SM prediction for the events in each of the 54 energy bins (27 bins for the near detector and 27 bins for the far detector) as well as the measured data, backgrounds and the covariance matrix.

The neutral current NSI cannot affect the charged current interaction rates so in our analysis, we merely focus on the neutral current events. Considering the stringent bounds on $\epsilon^{Vq}_{\alpha \beta}$ as well as  on  $\epsilon^{Aq}_{\mu \alpha }$, we mainly focus on  the $ee$, $e\tau$ and $\tau \tau$ components of $\epsilon^{Aq}$. For $\epsilon_{\mu \alpha}^{A/V q}=0$, the cross section of $\nu_\mu$ will not be affected. 
The energies of the bins of the combined  MINOS and MINOS+ data span from $\sim 1$ GeV to 40~GeV with the energy peaks at 3 GeV and 7 GeV respectively for MINOS and MINOS+. This rather  broad range covers quasi-elastic, resonance and Deep Inelastic Scattering (DIS) regimes. In \cite{Abbaslu:2023vqk,Abbaslu:2024hep},  we have discussed in detail how NSI affects the cross sections of these scatterings.  We use the NuWro Monte Carlo neutrino event generator to compute  the scattering cross section \cite{nuwro_github,Juszczak:2005zs,Golan:2012rfa,Golan:2012wx}.

\begin{figure}[htb!]
  \centering
  \includegraphics[width=0.78\textwidth]{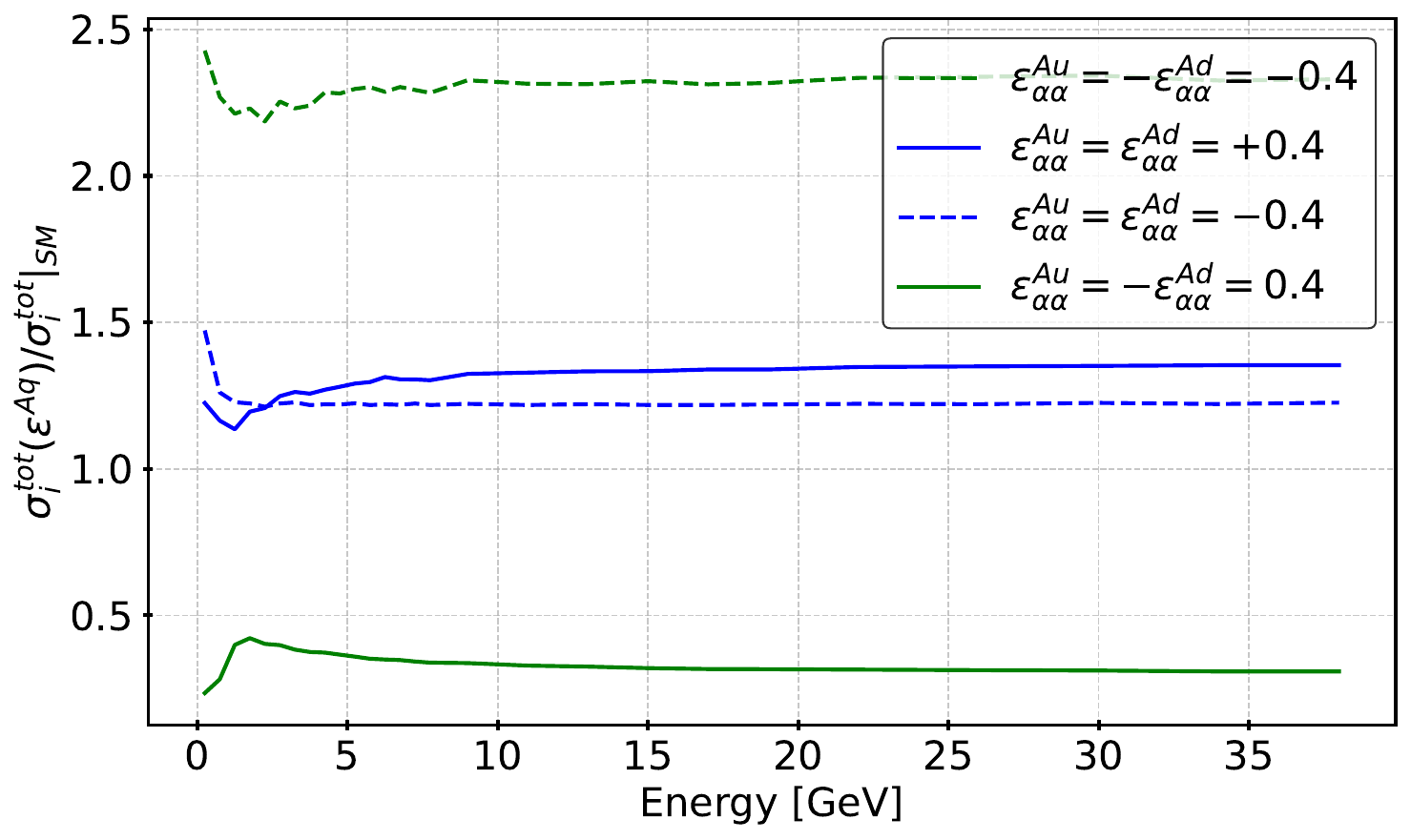}
  \caption{ The ratio of $\sigma^{tot}_i(\epsilon^{Aq})/\sigma^{tot}_i|_{SM}$ is shown for the isoscalar case, where $\epsilon^{Au}_{\alpha\alpha}=\epsilon^{Ad}_{\alpha\alpha}=\pm 0.4$ and the isovector case, where  $\epsilon^{Au}_{\alpha\alpha}=-\epsilon^{Ad}_{\alpha\alpha}=\pm 0.4$. The dashed green line represents the isovector case with $\epsilon^{Au}_{\alpha\alpha}=-\epsilon^{Ad}_{\alpha\alpha}=- 0.4$. The solid blue line shows the isoscalar case with $\epsilon^{Au}_{\alpha\alpha}=\epsilon^{Ad}_{\alpha\alpha}=0.4$. The dashed blue line corresponds to the isoscalar case $\epsilon^{Au}_{\alpha\alpha}=\epsilon^{Ad}_{\alpha\alpha}=-0.4$, and the solid green line corresponds to the isovector case $\epsilon^{Au}_{\alpha\alpha}=-\epsilon^{Ad}_{\alpha\alpha}=0.4$.}
  \label{fig:Ratio}
\end{figure}

\begin{figure}[htb!]
  \centering
  \includegraphics[width=0.787\textwidth]{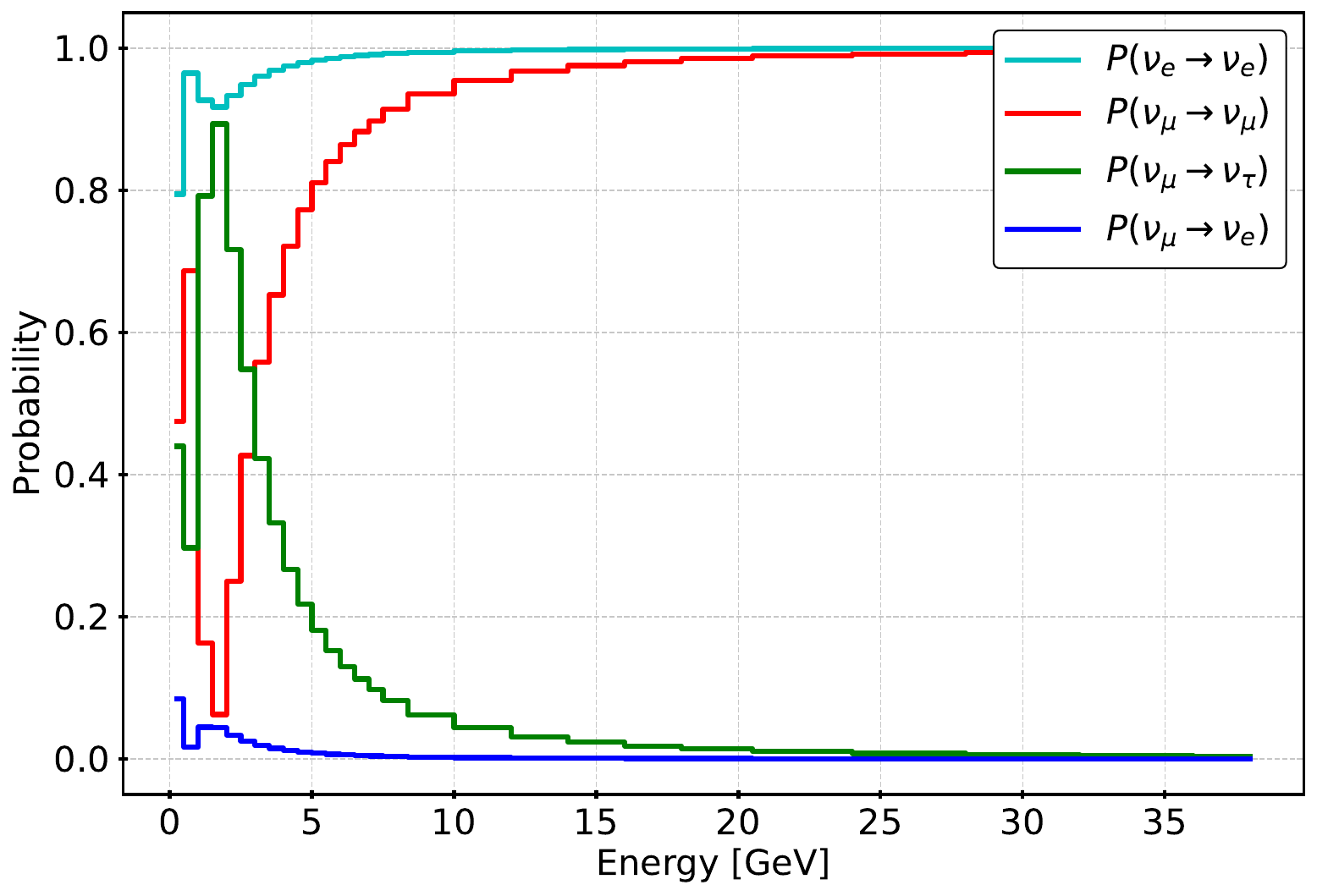}
  \caption{ The averaged neutrino oscillation probabilities per bin, as given by Eq.~\ref{Av-oscillation}, using the central values of oscillation parameters listed in table \ref{tab:mixing-parameters} for the MINOS and  MINOS$+$ experiments with a 735 km baseline.  }
  \label{fig:oscillation-probability}
\end{figure}

In the far detectors, the neutrino states are a linear combination of $\nu_e$, $\nu_\mu$ and $\nu_\tau$. Let us write
\begin{align}
	|\nu_{\rm far}(E_\nu)\rangle  \equiv \sum_\beta \mathcal{A}_\beta (E_\nu) |\nu_\beta (E_\nu)\rangle \quad,
\end{align}
in which $\mathcal{A}_\beta (E_\nu)$ is the amplitude of the oscillation $\nu_\mu \to \nu_\beta$ such that the oscillation probabilities are $$P(\nu_\mu \to \nu_\beta)=|\mathcal{A}_\beta (E_\nu)|^2.$$ 
Let us first focus on the lepton flavor conserving diagonal NSI. We will return to the case of off-diagonal lepton flavor violating NSI later on. If only the diagonal elements of $\epsilon^{Aq}$ or $\epsilon^{Vs}$ are nonzero, 
the interaction preserves lepton flavor so
\begin{align*} \sigma(\nu_{far}+{\rm Fe}\to \nu_\alpha +X)\propto |\mathcal{A}_\alpha (E_\nu)|^2\times |\mathcal{M}(\nu_\alpha +{\rm Fe} \to \nu_\alpha+X)|^2.
\end{align*}
Thus,  in the presence of diagonal NSI,
the number of the neutral current events at the $i$th  bin, $\mathcal{N}_i^{S}(\epsilon^{Aq})$ or $\mathcal{N}_i^{S}(\epsilon^{Vs})$, can be written in terms of that in the absence of NSI, $\mathcal{N}_i^{S}|_{SM}$ as 

\begin{align}
	\mathcal{N}_i^{S}(\epsilon)=\left(P_i(\nu_\mu \to \nu_e) \frac{\sigma^{tot}_i(\epsilon_{ee})}{\sigma^{tot}_i|_{SM}} + P_i(\nu_\mu \to \nu_\mu) \frac{\sigma^{tot}_i(\epsilon_{\mu\mu})}{\sigma^{tot}_i|_{SM}}\right.
\left.
    +P_i(\nu_\mu \to \nu_\tau) \frac{\sigma^{tot}_i(\epsilon_{\tau\tau})}{\sigma^{tot}_i|_{SM}}\right) \mathcal{N}_i^{S}|_{SM} \label{diag}
\end{align}
where $\sigma^{tot}_i|_{SM}$ is the total standard model (DIS +quasi-elastic+resonance scattering)  cross section  of the neutral current scattering of $\nu_\mu$ off nucleus. $\sigma^{tot}_i(\epsilon_{ee})$, $\sigma^{tot}_i(\epsilon_{\mu \mu})$ and $\sigma^{tot}_i(\epsilon_{\tau\tau})$ are  respectively the total cross sections of  the NC NSI scatterings of $\nu_e$, $\nu_\mu$ and $\nu_\tau$. In Eq. (\ref{diag}), we have neglected variation of the ratio $\sigma^{tot}_i(\epsilon^{Aq})/\sigma^{tot}_i|_{SM}$ within each energy bin.  Moreover, we have implicitly assumed that this ratio at the true neutrino energy and the reconstructed neutrino energy are equal. As seen in Fig~\ \ref{fig:Ratio}, this ratio is almost constant when we vary the energy so these approximations do not induce any significant error.
\footnote{It is worth mentioning that for higher energies where Deep Inelastic Scattering (DIS) is dominant, the variation of the ratio with the energy is completely negligible. For $E_\nu<4$~GeV,  the variation mostly comes from the transition from the quasi-elastic regime to the resonance and then to the DIS regime.}
 $P_i(\nu_\alpha \to \nu_\beta)$ are the average oscillation probability for each bin. As explained before as long as the nonzero NSI couplings are $\epsilon^{Aq}$ and $\epsilon^{Vs}$, the oscillation is not affected by NSI.  Unlike 
 $\sigma_i^{tot}(\epsilon)/\sigma_i^{tot}|_{SM}$, the oscillation probabilities can dramatically vary for $E_\nu<2$~GeV such that over the low energy bins with a width of 0.5~GeV, $P(\nu_\mu \to \nu_\alpha)$ can significantly change. To account for this variation, we have defined
 \begin{equation}\label{Av-oscillation} 
P_i(\nu_\alpha \to \nu_\beta)\equiv \frac{\int_{E_{min}^i}^{E_{max}^i} P_i(\nu_\alpha \to \nu_\beta) dE}{E_{max}^i-E_{min}^i} 
\end{equation}
in which $E_{max}^i$ and $E_{min}^i$ are the limits of the $i$th bin. $P_i(\nu_\alpha \to \nu_\beta)$ are shown in Fig. \ref{fig:oscillation-probability} where the central values of the neutrino mass and mixing parameters shown in Tab \ref{tab:mixing-parameters} are taken as the input. Notice that in Eq. (\ref{Av-oscillation}), it is implicitly assumed that the flux is constant over each bin. Indeed, 
the variation of the flux over each bin, $\Delta F_i/F_i$, is less  than about 10 percent (see Refs. \cite{Bishai:2012kta,Wood:2024jos})  so this assumption is justified. On the other hand, the variation of oscillation over the bin size is $\Delta P_i/P_i \stackrel{<}{\sim} 2(E_{max}^i-E_{min}^i)/(E_{max}^i+E_{min}^i).$ Thus, the error induced by taking the flux, $F_i$ constant over a bin is given by $(\Delta F_i/F_i)(\Delta P_i/P_i)\sim$ few percent which is negligible compared to the effects of the NSI couplings $\epsilon_{\tau (e) \tau (e)}$.

At ND, $P_i(\nu_\mu \to \nu_e)=P_i(\nu_\mu \to \nu_\tau)=0$ and $P_i(\nu_\mu \to \nu_\mu)=1$ so as long as $\epsilon_{\mu\mu}=0$, $\mathcal{N}_i^S$ for the ND bins should be given by the SM prediction. However, including the ND bins in the analysis ({\it i.e.,} in the definition of $\chi_{NC}^2$ in Eq. (\ref{ChiNC})) reduces the correlated uncertainties such as those originating from the flux normalization. As discussed in sect. \ref{sec:Bounds}, there are already stringent bounds on $\epsilon_{\mu \alpha}^{A/V u/d}$. However, $\epsilon_{\mu \alpha}^{As}$  and  $\epsilon_{\mu \alpha}^{Vs}$  can be large. 

In the presence of nonzero $\epsilon^{Aq}_{e\tau}$, the amplitudes of the scattering of $\nu_{\rm far}$  to $\nu_e$ and $\nu_\tau$ from the different flavor components of $|\nu_{\rm far}(E_\nu)\rangle$ can interfere so we cannot use the simple formula in Eq. (\ref{diag}). We take $\epsilon_{e\tau}^{Aq}$ to be a real parameter which can be negative, positive or zero. We denote the phase of the $e \tau$ component with $\alpha \in  (0,\pi)$. Let us  change the basis from $(\nu_e,\nu_\mu,\nu_\tau)$ to  $(\tilde{\nu}_e,\nu_\mu,\tilde{\nu}_\tau)$ in which the neutral current couplings are diagonal:

\begin{align}
	\frac{G_F}{\sqrt{2}}(\bar{q}\gamma^\mu \gamma^5 q)[ \bar{\nu}_e \ \bar{\nu}_\mu \ \bar{\nu}_\tau]  \times\left[ 
	\begin{matrix}  g^{Aq} +\epsilon^{Aq}_{ee}  & 0 & \epsilon^{Aq}_{e\tau}e^{-i \alpha} \cr 0 &  g^{Aq} & 0 \cr  \epsilon^{Aq}_{e\tau} e^{i \alpha} &0 & g^{Aq} +\epsilon^{Aq}_{\tau \tau}  \end{matrix} 
	\right]\gamma_\mu (1-\gamma^5)
	\left[ \begin{matrix} \nu_e \cr \nu_\mu \cr \nu_\tau \end{matrix} \right]=
\end{align}

\begin{align}
	\frac{G_F}{\sqrt{2}}(\bar{q}\gamma^\mu \gamma^5 q)[ \bar{\tilde{\nu}}_e \ \bar{\tilde{\nu}}_\mu \ \bar{\tilde{\nu}}_\tau] \times\left[ 
	\begin{matrix}  g^{Aq} +\tilde{\epsilon}^{Aq}_{ee}  & 0 & 0\cr 0 &  g^{Aq} & 0 \cr  0  &0 & g^{Aq} +\tilde{\epsilon}^{Aq}_{\tau \tau}  \end{matrix} 
	\right] \gamma_\mu (1-\gamma^5)
	\left[ \begin{matrix} \tilde{\nu}_e \cr  \tilde{\nu}_\mu \cr  \tilde{\nu}_\tau \end{matrix} \right] \nonumber
\end{align}
in which $g^{Au}= \frac{1}{2}$ and $g^{Ad}=-\frac{1}{2}$ are the axial neutral current couplings within the standard model and
\begin{equation}
	\left[ \begin{matrix} \tilde{\nu}_e \cr  \tilde{\nu}_\mu \cr  \tilde{\nu}_\tau \end{matrix} \right]=\left[\begin{matrix} \cos \theta e^{i\alpha} & 0 & \sin \theta \cr 0 &1& 0 \cr-\sin \theta e^{i\alpha}& 0 & \cos \theta \end{matrix}\right] 	\left[ \begin{matrix} \nu_e \cr \nu_\mu \cr \nu_\tau \end{matrix} \right] \  {\rm where} \ \tan 2 \theta= \frac{2 \epsilon_{e\tau}^{Aq}}{ \epsilon_{ee}^{Aq}-\epsilon_{\tau\tau}^{Aq}}
\end{equation}
and 
\begin{align*}
&\tilde{\epsilon}^{Aq}_{ee}= \frac{\epsilon^{Aq}_{ee} + \epsilon^{Aq}_{\tau\tau}}{2} + \frac{1}{2}\sqrt{(\epsilon^{Aq}_{ee} - \epsilon^{Aq}_{\tau\tau})^2 + 4 (\epsilon^{Aq}_{e\tau})^2} \ \ {\rm and} \ \\ &\tilde{\epsilon}^{Aq}_{\tau\tau}= \frac{\epsilon^{Aq}_{ee} + \epsilon^{Aq}_{\tau\tau}}{2} -\frac{1}{2} \sqrt{(\epsilon^{Aq}_{ee} - \epsilon^{Aq}_{\tau\tau})^2 + 4 (\epsilon^{Aq}_{e\tau})^2}.\end{align*}
For a general value of $\epsilon^{Au}/\epsilon^{Ad}$, we cannot find a single basis in which the couplings to the $u$ and $d$ quarks are both diagonal. However, for the particular cases that we consider here ({\it i.e.,} $\epsilon^{Au}=\pm \epsilon^{Ad}$ or  $\epsilon^{Au}=0$ or $ \epsilon^{Ad}=0$),  a single  mixing matrix  can diagonalize the couplings to  both the $u$ and $d$ quarks. 
We can then compute $\mathcal{N}_i^{FD}(\epsilon^{Aq})$ with Eq. (\ref{diag}) replacing $\epsilon^{Aq}_{\alpha\alpha} \to \tilde{\epsilon}^{Aq}_{\alpha\alpha}$ and $P(\nu_\mu \to \nu_\alpha )\to  P(\nu_\mu \to \tilde{\nu}_\alpha )$ in which
\begin{eqnarray}
	P_i(\nu_\mu \to \tilde{\nu}_e) &= &\frac{\int_{E^i_{min}}^ {E^i_{max}}|\cos \theta \mathcal{A}_e e^{-i\alpha}+\sin \theta \mathcal{A}_\tau |^2dE_\nu}{E_{max}^i-E_{min}^i} 
	\cr
	P_i(\nu_\mu \to \tilde{\nu}_\mu) &= &P_i(\nu_\mu \to \nu_\mu)=\frac{\int_{E^i_{min}}^ {E^i_{max}}| \mathcal{A}_\mu |^2 dE_\nu}{E_{max}^i-E_{min}^i}\cr
	P_i(\nu_\mu \to \tilde{\nu}_\tau) &=&\frac{\int_{E^i_{min}}^ {E^i_{max}}|-\sin \theta  \mathcal{A}_e e^{-i\alpha}+\cos \theta \mathcal{A}_\tau |^2 dE_\nu}{E_{max}^i-E_{min}^i} \ .
\end{eqnarray}
Notice that $\epsilon^{Aq}$ or $\epsilon^{Vs}$ do not change the oscillation amplitude, $\mathcal{A}_\alpha$. If we considered $\epsilon^{Vu}$ and $\epsilon^{Vd}$, the amplitude would change and we should have used the formulas in the Appendix of \cite{Coloma:2022umy}.

For $\epsilon_{e\tau}^{Aq}=0$ but $\epsilon_{\mu\tau}^{Aq}\ne 0$, we can make a similar basis change, $(\nu_e,\nu_\mu,\nu_\tau) \to (\hat{\nu}_e,\hat{\nu}_\mu,\hat{\nu}_\tau) $
with $\hat{\nu}_e=\nu_e$ and mixing between the  $\mu $
 and $\tau$ neutrinos given by $\tan 2 \theta=2 \epsilon_{\mu\tau}^{Aq}/(\epsilon^{Aq}_{\mu \mu}-\epsilon^{Aq}_{\tau\tau})$ and therefore
\begin{eqnarray}
	P_i(\nu_\mu \to \hat{\nu}_e) &= &P_i(\nu_\mu \to {\nu}_e)=\frac{\int_{E^i_{min}}^ {E^i_{max}}| \mathcal{A}_e |^2dE_\nu}{E_{max}^i-E_{min}^i} 
	\cr
	P_i(\nu_\mu \to \hat{\nu}_\mu) &= &\frac{\int_{E^i_{min}}^ {E^i_{max}}| \cos \theta\mathcal{A}_\mu e^{-i\beta} +\sin \theta\mathcal{A}_\tau|^2 dE_\nu}{E_{max}^i-E_{min}^i}\cr
	P_i(\nu_\mu \to \hat{\nu}_\tau) &=&\frac{\int_{E^i_{min}}^ {E^i_{max}}|-\sin \theta \mathcal{A}_\mu e^{-i\beta}+\cos \theta \mathcal{A}_\tau |^2 dE_\nu}{E_{max}^i-E_{min}^i} \ ,
\end{eqnarray}
 where $\beta$ is the  phase of the ${\mu\tau}$ component. 

\begin{figure*}[t!]
    \centering
    \subfigure[]{\includegraphics[width=0.48\textwidth ]{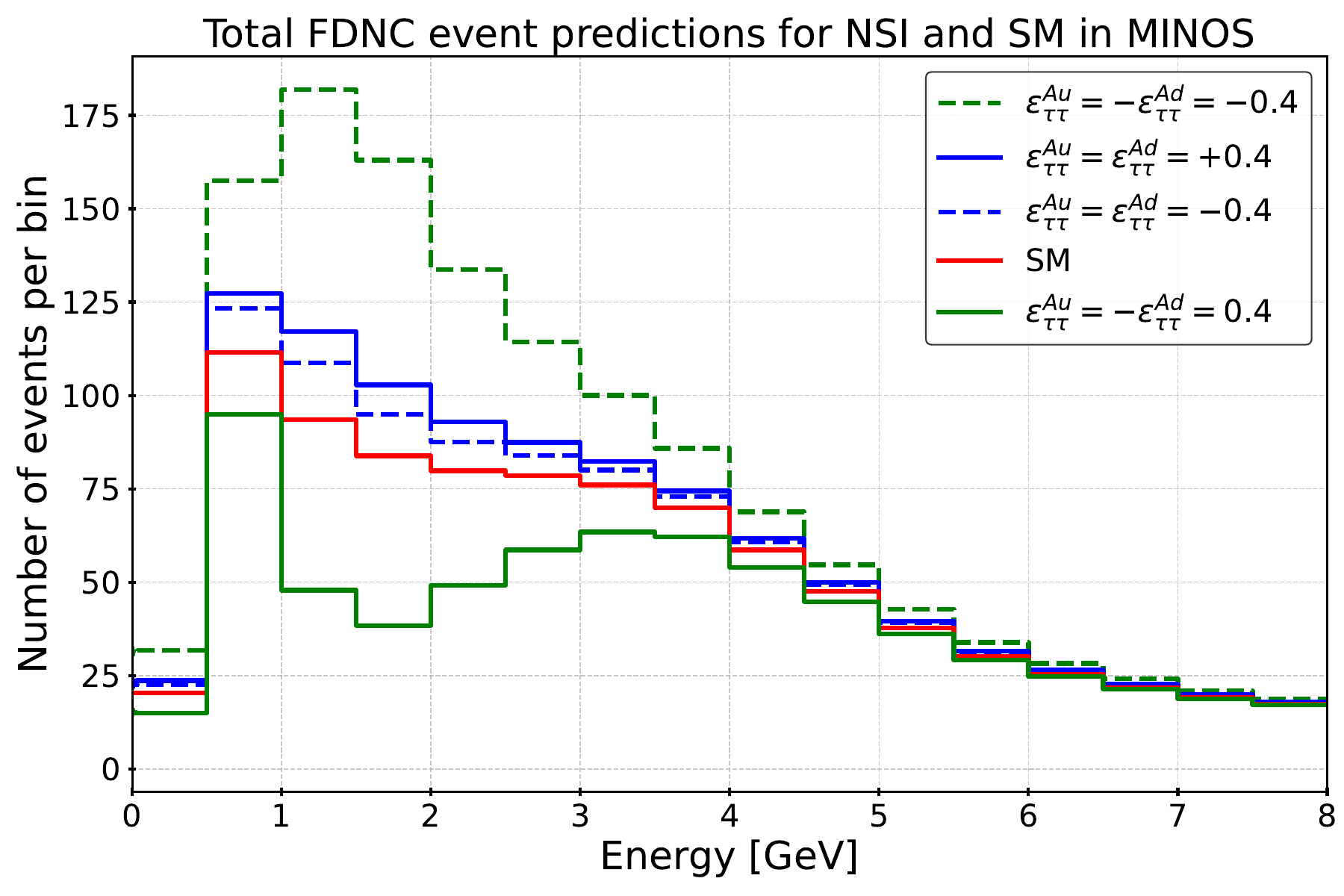}\label{FDNC_prediction_NSI-minos-a}}
    \subfigure[]{\includegraphics[width=0.48\textwidth ]{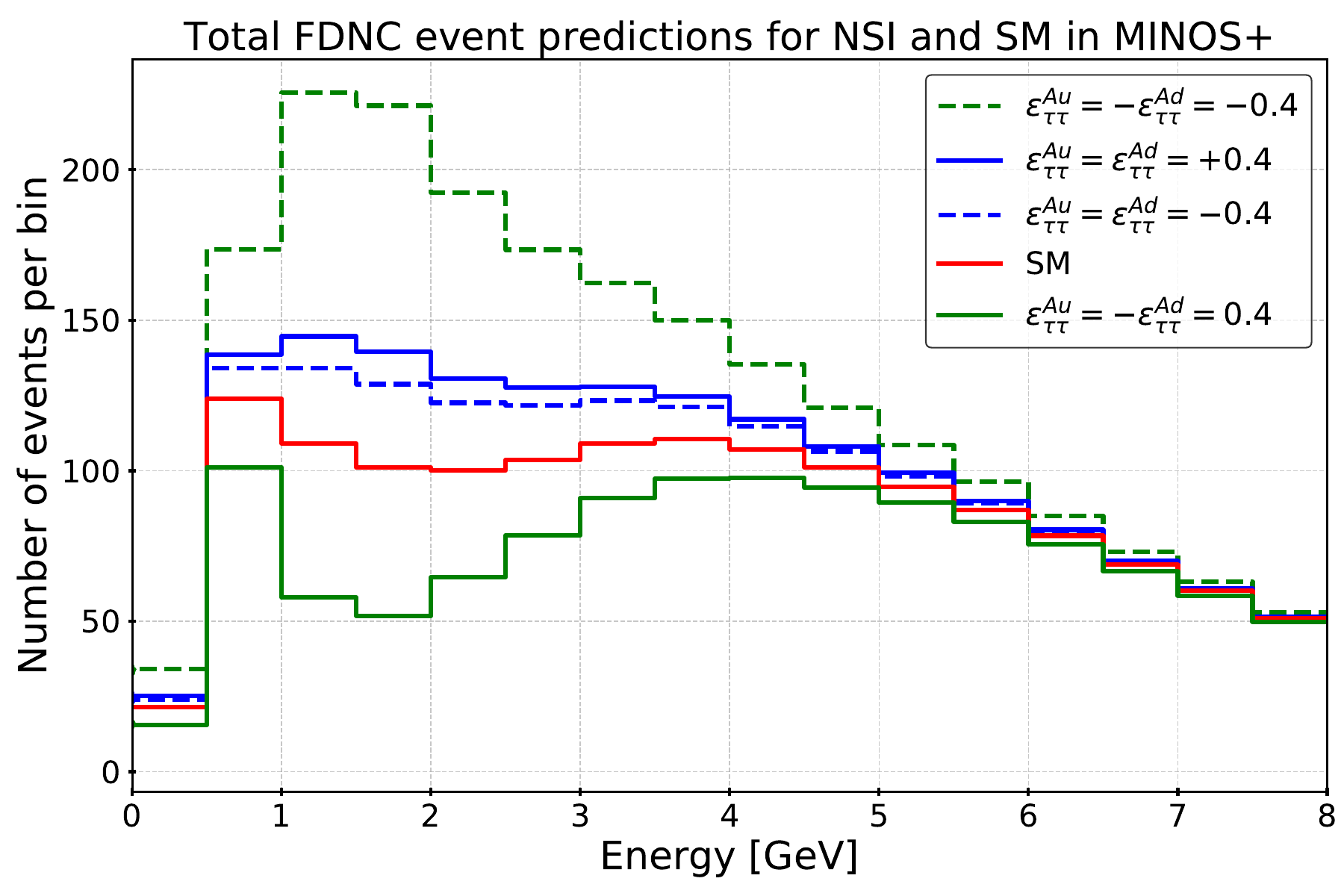}
        \label{FDNC_prediction_NSI-minosplus-b}}
    \vspace{-0.8em}
    \caption{Total number of predicted  NC events per bin (signal plus all backgrounds) at the FD, with and without Non-Standard Interactions (NSI). The dashed green histograms show the total number of events   for $\epsilon^{Au}_{\tau\tau}=-\epsilon^{Ad}_{\tau\tau}=-0.4$. The solid blue histograms show  the total number of events   for $\epsilon^{Au}_{\tau\tau}=\epsilon^{Ad}_{\tau\tau}=0.4$. The dashed blue histograms show the total number of events   for $\epsilon^{Au}_{\tau\tau}=\epsilon^{Ad}_{\tau\tau}=-0.4$. The solid red histograms show the total number of events for SM case {\it i.e.,} $\epsilon^{Au}_{\tau\tau}=\epsilon^{Ad}_{\tau\tau}=0$. The solid green histograms show the total number of events for $\epsilon^{Au}_{\tau\tau}=-\epsilon^{Ad}_{\tau\tau}=0.4$. The left (right) panels are the predictions for MINOS (MINOS+). The width of the shown bins is 0.5 GeV.}\label{fig:FD-Minos-NSISM}
\end{figure*}

Figure \ref{fig:FD-Minos-NSISM} shows the predicted number of NC events plus the background per bins of 0.5 GeV for the SM (in red) and various values of the NSI coupling at the MINOS and MINOS+ far detectors. As seen in the figure, the deviation from the SM prediction can be sizable.  This promises the possibility of constraining $\epsilon^{Aq}$.  We find that the MINOS and MINOS+ NC data is consistent with the SM prediction with $\epsilon^{Aq}=0$. In the next section, we show the bounds from the MINOS and MINOS+ NC data that we have derived. Notice that, unlike the oscillation to the sterile neutrinos, NSI can lead to both excess and deficit of the NC events. If in the future measurements at the  DUNE far detector, an excess of NC events is  observed, it can be interpreted as a hint for the axial NSI rather than $\nu_\alpha \to \nu_s$ or other scenarios that predict only deficit.

\section{Bounds from MINOS and MINOS+ on the axial NSI couplings \label{sec:OURS}}

First in this section, we focus on the case of  one nonzero NSI coupling, setting the rest of the NSI couplings to zero. Various flavor symmetries can impose such a flavor structure 
(see, for example, Ref. \cite{Abbaslu:2024hep} which predicts $\epsilon_{\tau \tau}^{Au}=\epsilon^{Ad}_{\tau\tau} \ne  0$ and $\epsilon_{\alpha \tau}=\epsilon_{\tau\alpha }=\epsilon_{\alpha \beta}=0$ with $\alpha,\beta \in \{ e, \mu\}$).  We will then relax this assumption and will study the correlation effects among various NSI couplings.
As discussed in the previous section, since the NSI in general has a non-universal flavor structure, the number of neutral current events at the far detector of MINOS will depend on the oscillation probabilities (see Eq. (\ref{diag})). At the far detector $\Delta m_{21}^2L/E_\nu\ll \Delta m_{31}^2L/E_\nu \sim 1$ so the oscillation probability will be sensitive to $\theta_{23}$, $\theta_{13}$, $\delta$ and $\Delta m_{31}^2$ but the sensitivity to $\theta_{12}$ and $\Delta m_{21}^2$ will be insignificant. The uncertainties in $\theta_{23}$, $\theta_{13}$, $\delta$ and $\Delta m_{31}^2$ can therefore affect probing the NSI couplings and should be taken into account when deriving bounds on NSI.

To explore the multidimensional parameter space of standard neutrino oscillation parameters and non-standard interaction (NSI) parameters, we perform a Markov Chain Monte Carlo (MCMC) analysis using the
Cobaya framework with a Metropolis-Hastings sampler. The log-likelihood function can be defined as $\mathcal{L} = -0.5 \, \chi_{NC}^2,$ where $\chi_{NC}^2$ is given in Eq. (\ref{ChiNC}).
The parameter space includes the standard three-flavor oscillation parameters $\Delta m_{31}^2$, $\theta_{23}$, $\theta_{13}$, and $\delta$, along with the NSI parameters $\epsilon_{ee}^{Aq}$, $\epsilon_{\mu\mu}^{Aq}$, $\epsilon_{\tau\tau}^{Aq}$, $\epsilon_{e\tau}^{Aq}$, and the complex phase $\alpha$. Since the sensitivity to $\theta_{12}$  and $\Delta m_{21}^2$ is negligible, we fix these parameters to their global best fit values in Table \ref{tab:mixing-parameters}. We then apply
Gaussian priors to $\Delta m_{31}^2$, $\theta_{23}$, $\theta_{13}$ and $\delta$  with central values and widths given in Table  \ref{tab:mixing-parameters}. Similarly, we apply the bound on   $\epsilon_{\mu\mu}^{Aq}$   using a Gaussian prior with $\mathcal{N}(0, 0.01)$. For $\epsilon_{ee}^{Aq}$, $\epsilon_{\tau\tau}^{Aq}$, $\epsilon_{e\tau}^{Aq}$ and $\alpha$, we apply uniform priors over the specified physical ranges. 
We quantify the convergence using the Gelman–Rubin diagnostic with stopping criterion $ R-1 < 0.001$  or a maximum steps of $10^6$.
We allow 
sampler iteration  until  reaching convergence. Moreover, we marginalize the posterior distributions to produce one-dimensional constraints on single NSI parameters.

\begin{figure*}[htb]
    \centering
    \subfigure[]{\includegraphics[width=0.48\textwidth ]{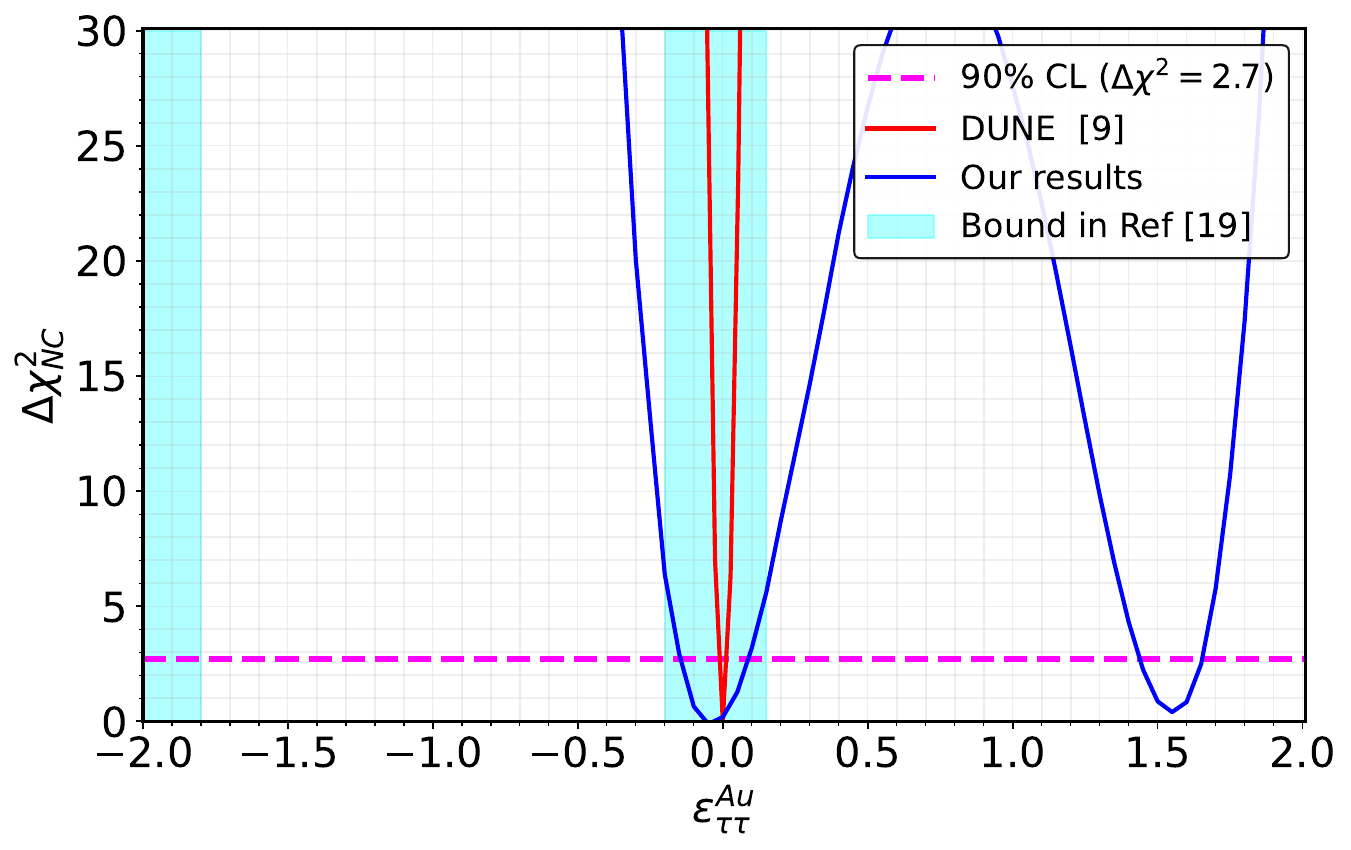}\label{fig:delta-chi2-tt-Au}}
    \subfigure[]{\includegraphics[width=0.48\textwidth ]{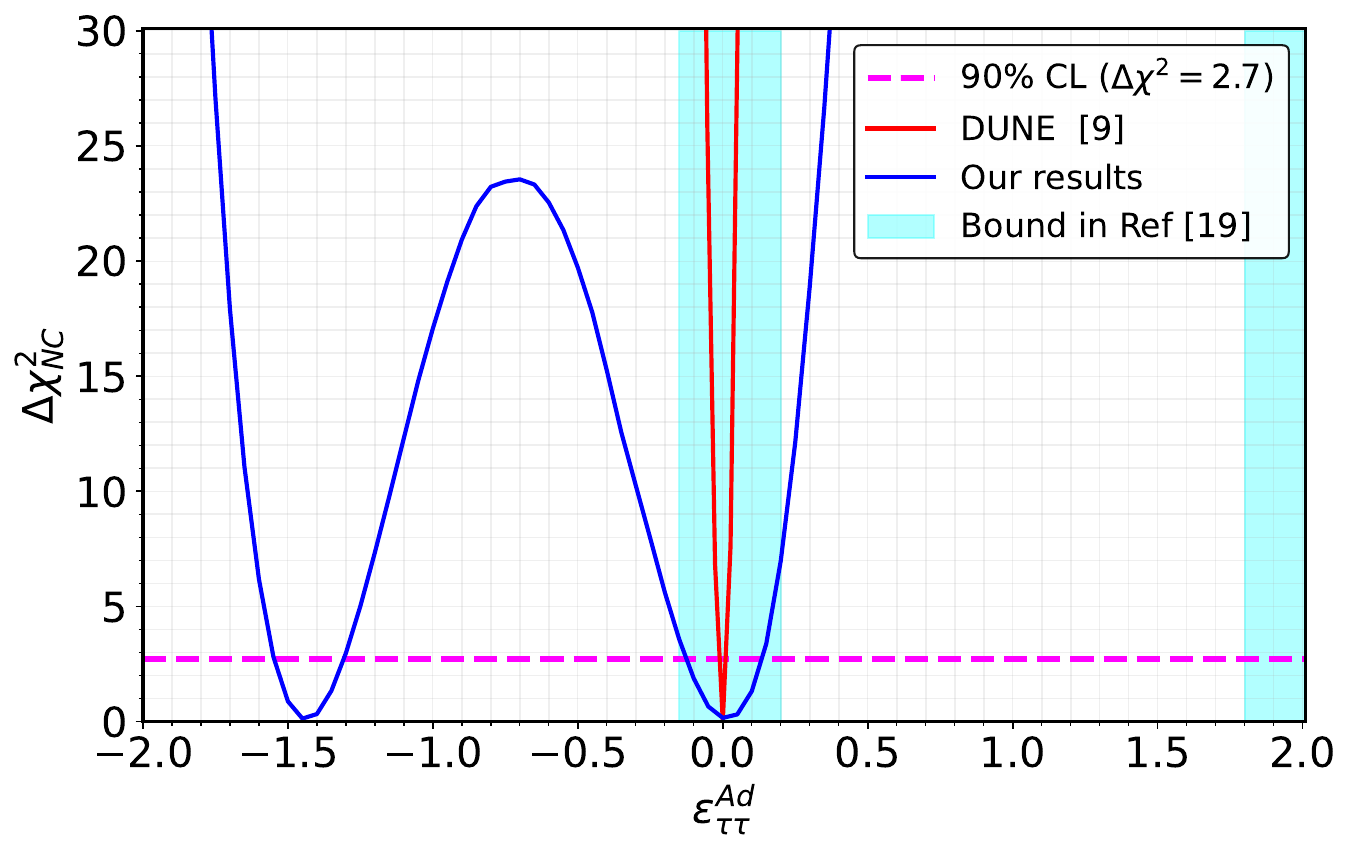}\label{fig:delta-chi2-tt-Ad}}
    \subfigure[]{\includegraphics[width=0.48\textwidth ]{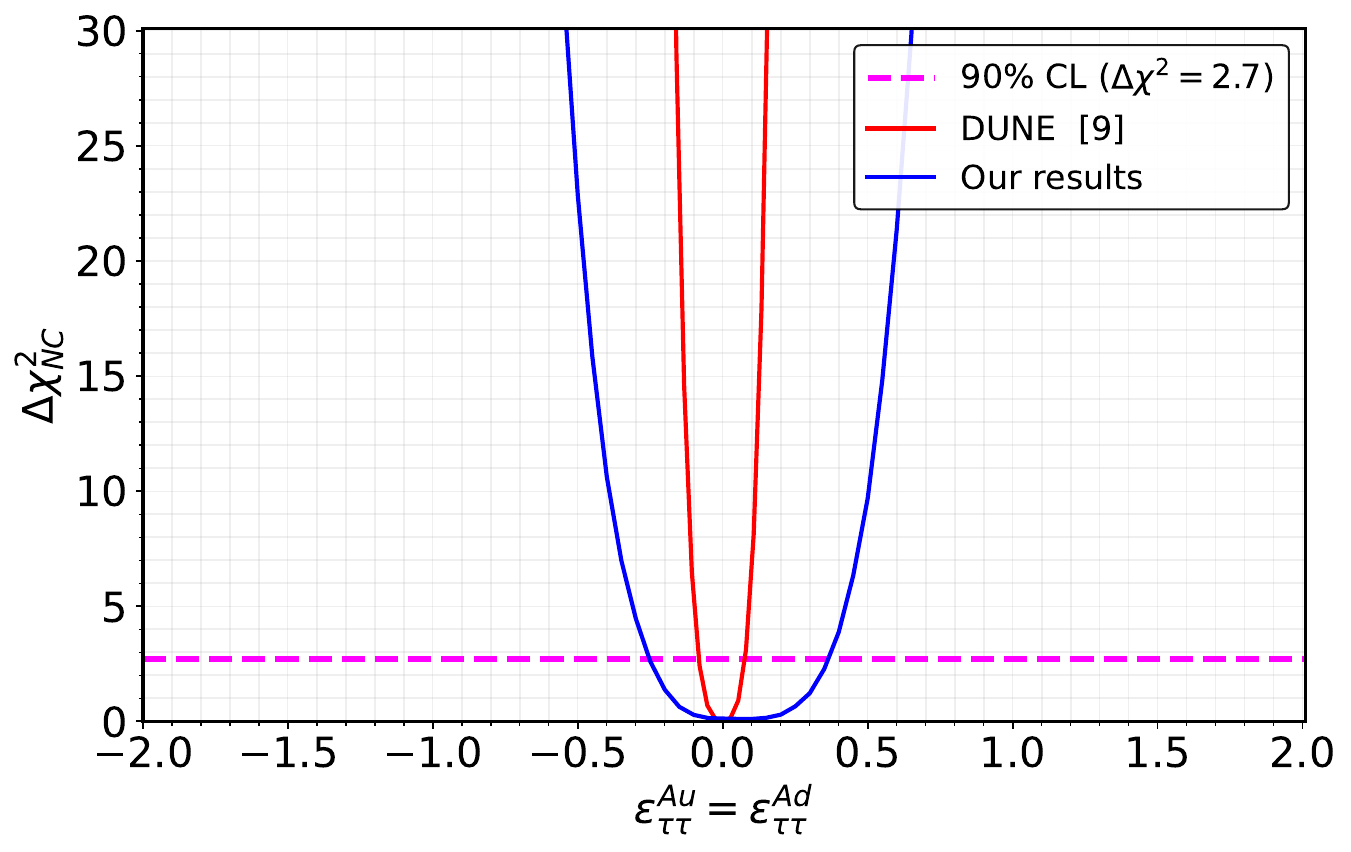}\label{fig:delta-chi2-tt-Au=Ad}}
    \subfigure[]{\includegraphics[width=0.48\textwidth ]{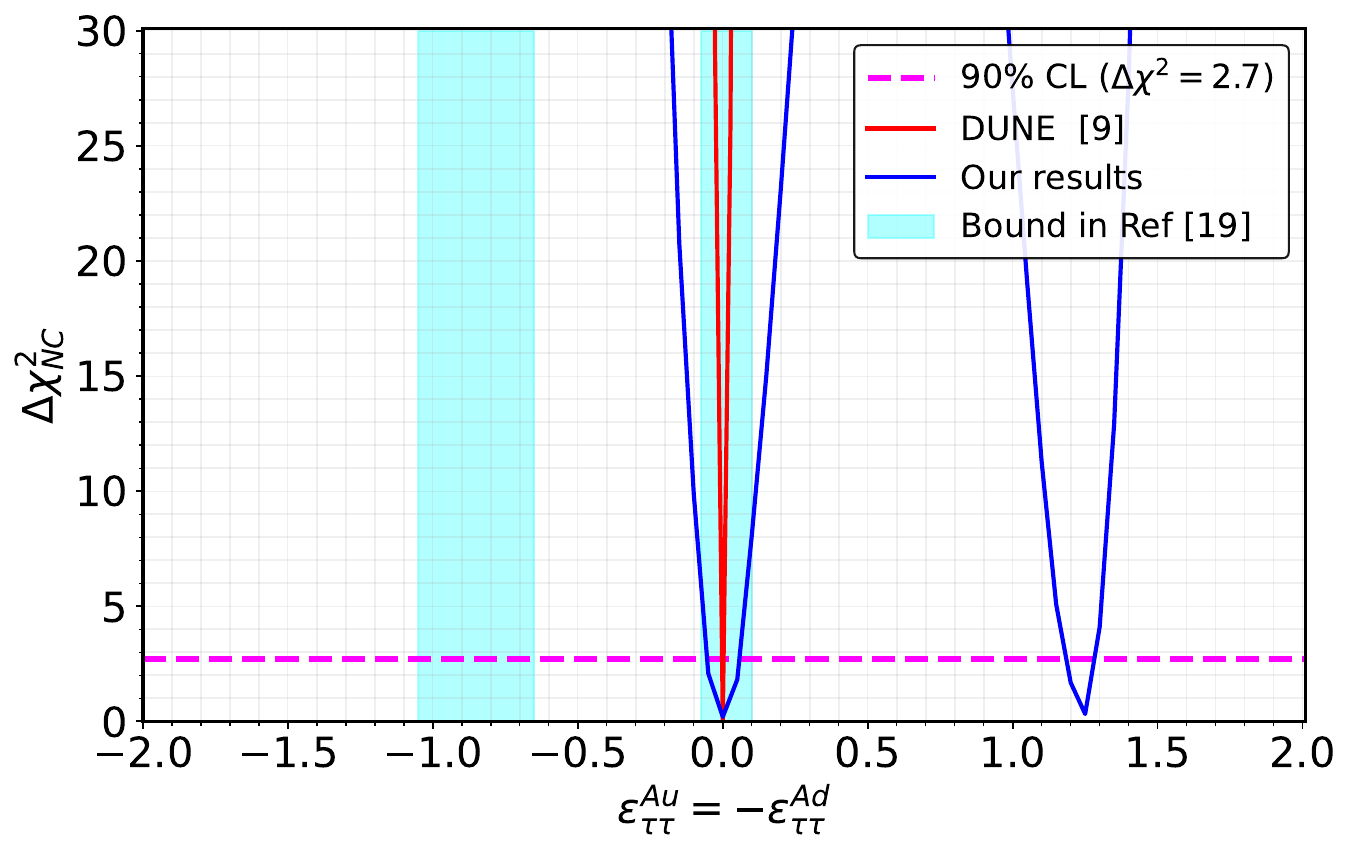}\label{fig:delta-chi2-tt-Au=-Ad}}
    \caption{ $\Delta\chi^2_{NC}$ distributions for different ratios of $\epsilon^{Au}_{\tau\tau}/\epsilon^{Ad}_{\tau\tau}$. 
    The label of the horizontal axis in each panel shows the nonzero NSI
    coupling. The rest of the NSI couplings are set to zero.
   The horizontal magenta dashed lines show $\Delta \chi^2_{NC}=2.7$, corresponding to 90 \% C.L. with one degrees of freedom. The red lines show the ideal bounds forecast for the DUNE experiment, assuming negligible systematic error \cite{Abbaslu:2023vqk}.  The blue curves show our results based on the MINOS and MINOS+ data.  We have marginalized over the relevant oscillation parameters using a 
Gaussian priors with widths given by the $1\sigma$ confidence intervals   as shown in Table \ref{tab:mixing-parameters}. The cyan bands correspond to the previous 90 \% C.L. bounds \cite{Coloma:2023ixt} in the literature, dominated by the SNO NC data. }
    \label{fig:delta-chi2-tt}
\end{figure*}

\begin{figure*}[]
    \centering
    \subfigure[]{\includegraphics[width=0.42\textwidth ]{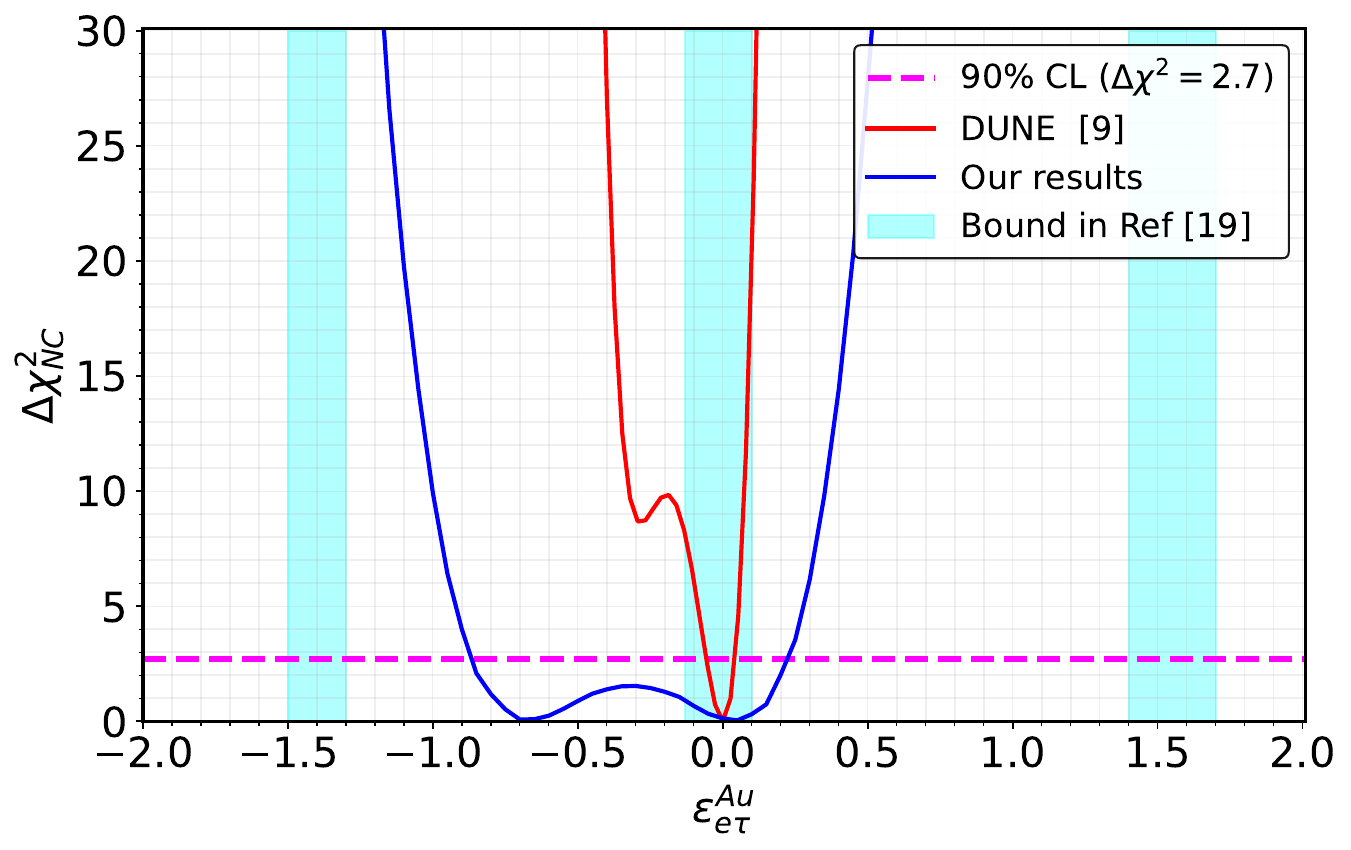}\label{fig:delta-chi2-et-Au}}
    \subfigure[]{\includegraphics[width=0.42\textwidth ]{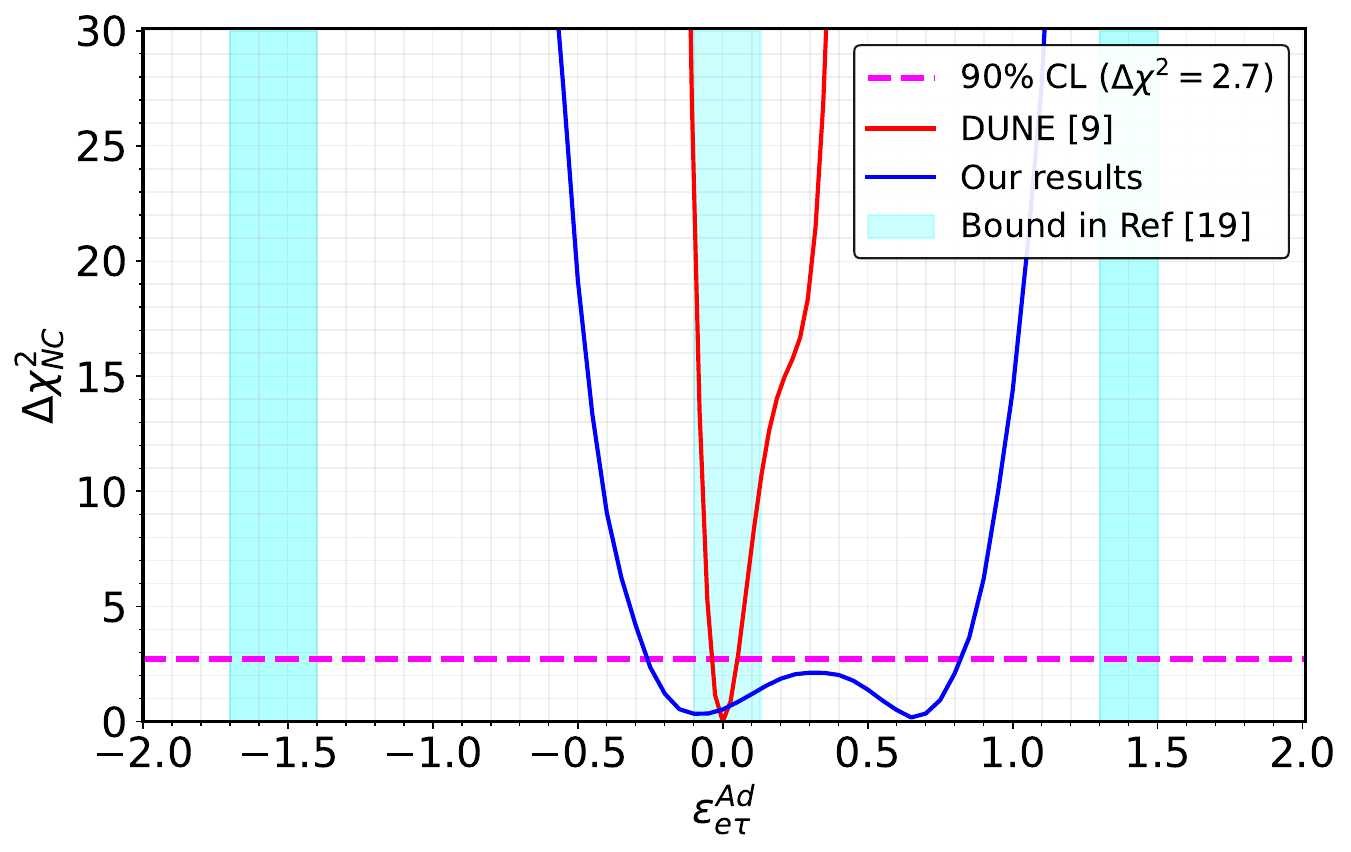}\label{fig:delta-chi2-et-Ad}}
    \subfigure[]{\includegraphics[width=0.42\textwidth ]{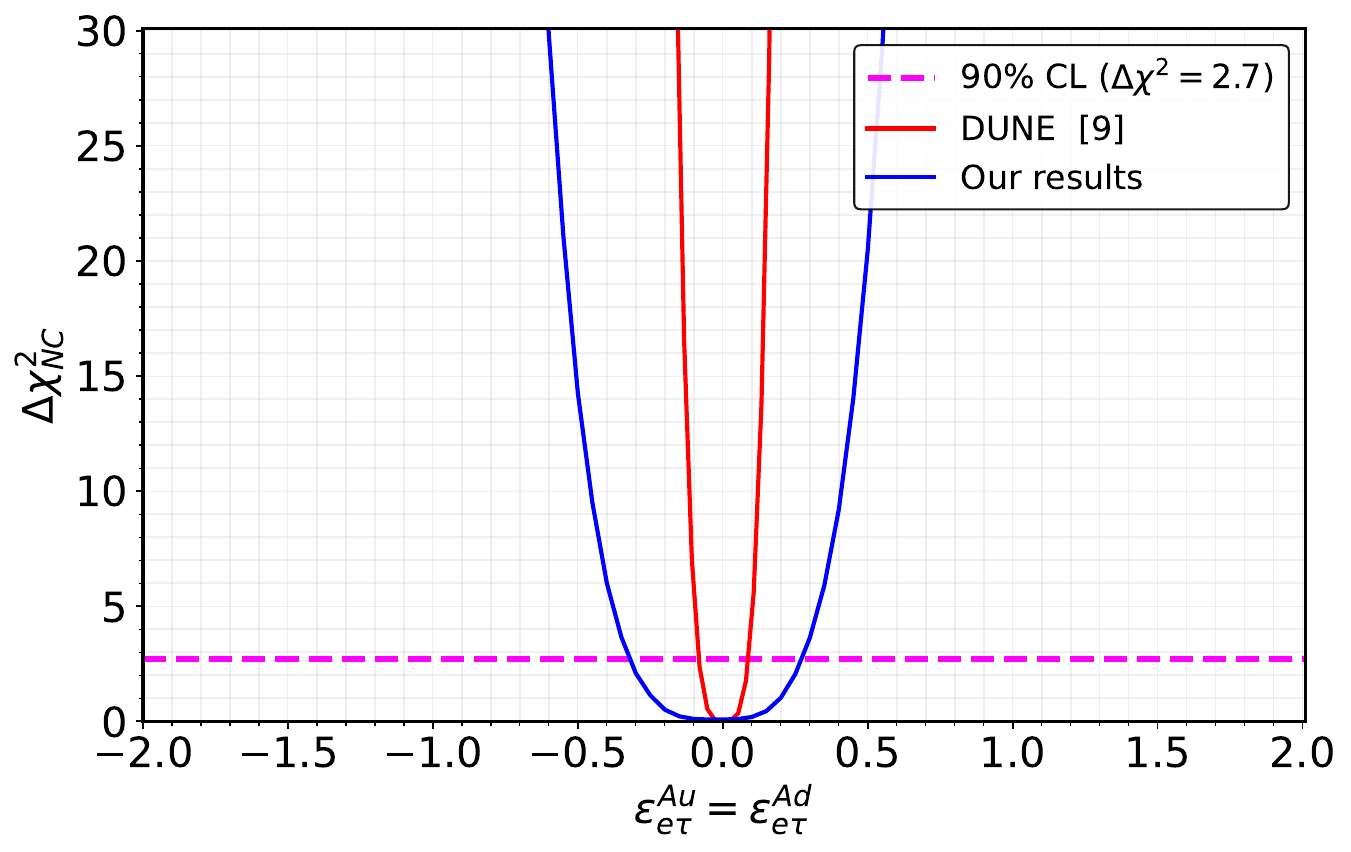}\label{fig:delta-chi2-et-Au=Ad}}
    \subfigure[]{\includegraphics[width=0.42\textwidth ]{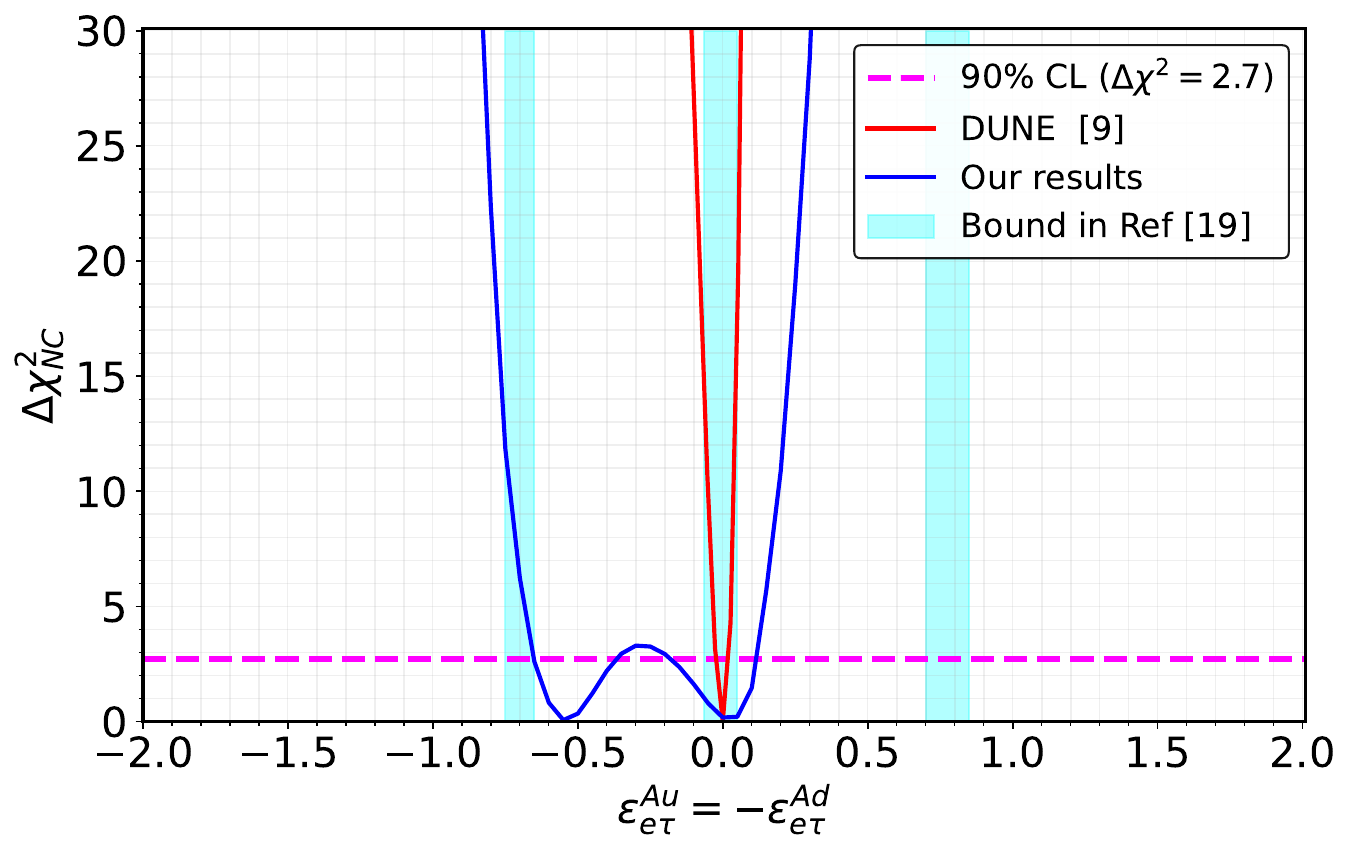}\label{fig:delta-chi2-et-Au=-Ad}}
    \caption{ $\Delta\chi^2_{NC}$ distributions for different $\epsilon^{Au}_{e\tau}/\epsilon^{Ad}_{e\tau}$ ratios. See the caption of  Fig. \ref{fig:delta-chi2-tt} for further explanation.}
    \label{fig:delta-chi2-et}
\end{figure*}

\begin{figure*}[]
    \centering
        \subfigure[]{\includegraphics[width=0.42\textwidth ]{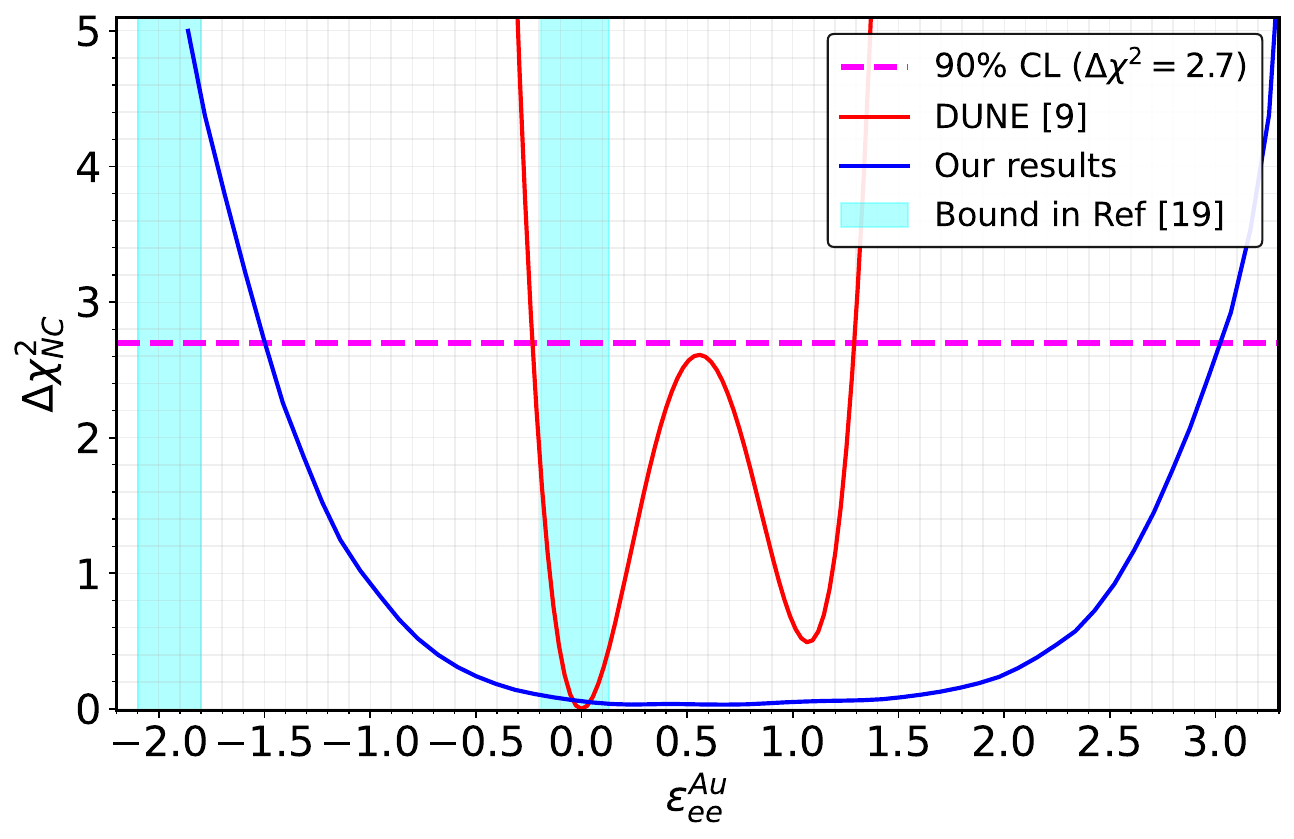}\label{fig:delta-chi2-ee-Au}}
        \subfigure[]{\includegraphics[width=0.42\textwidth ]{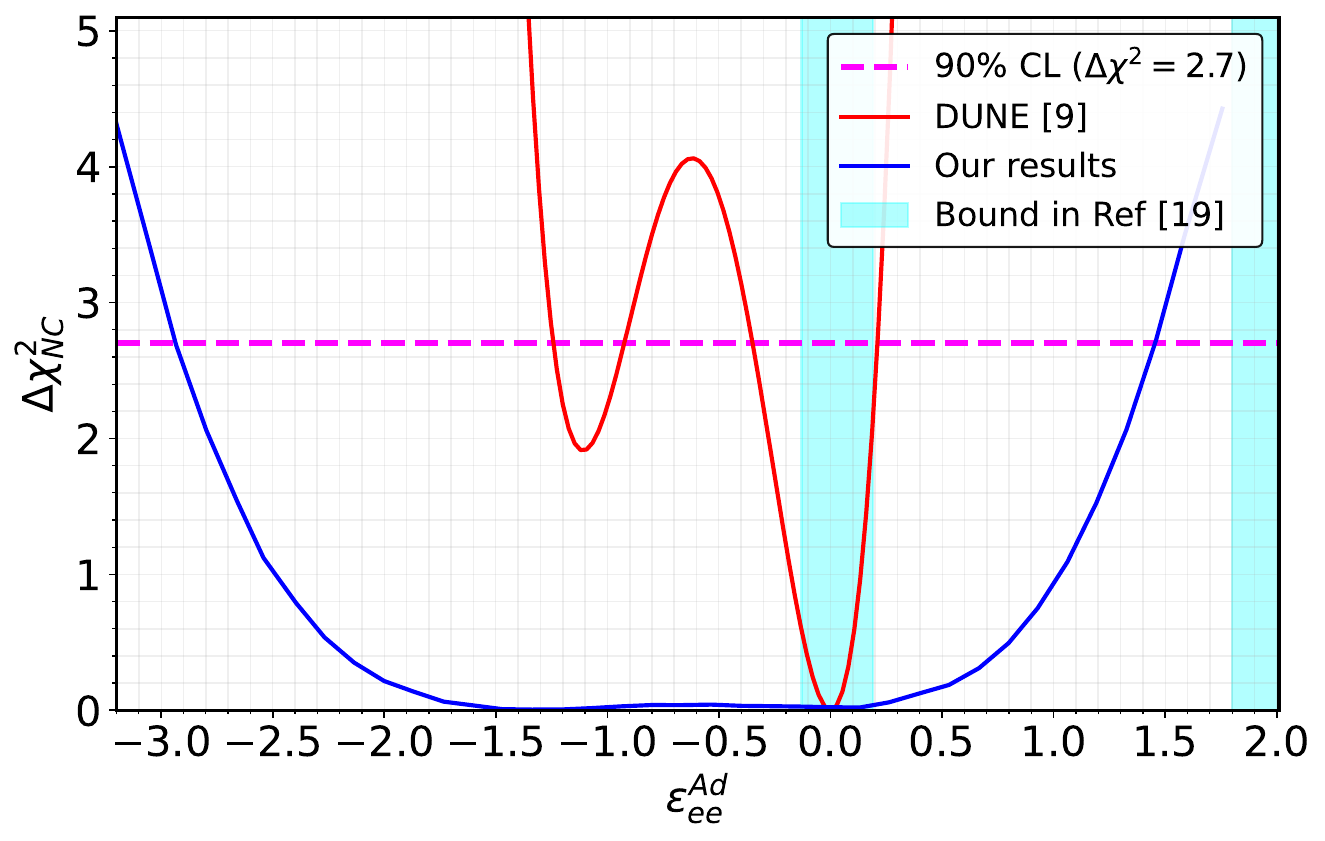}\label{fig:delta-chi2-ee-Ad}}
    \subfigure[]{\includegraphics[width=0.42\textwidth ]{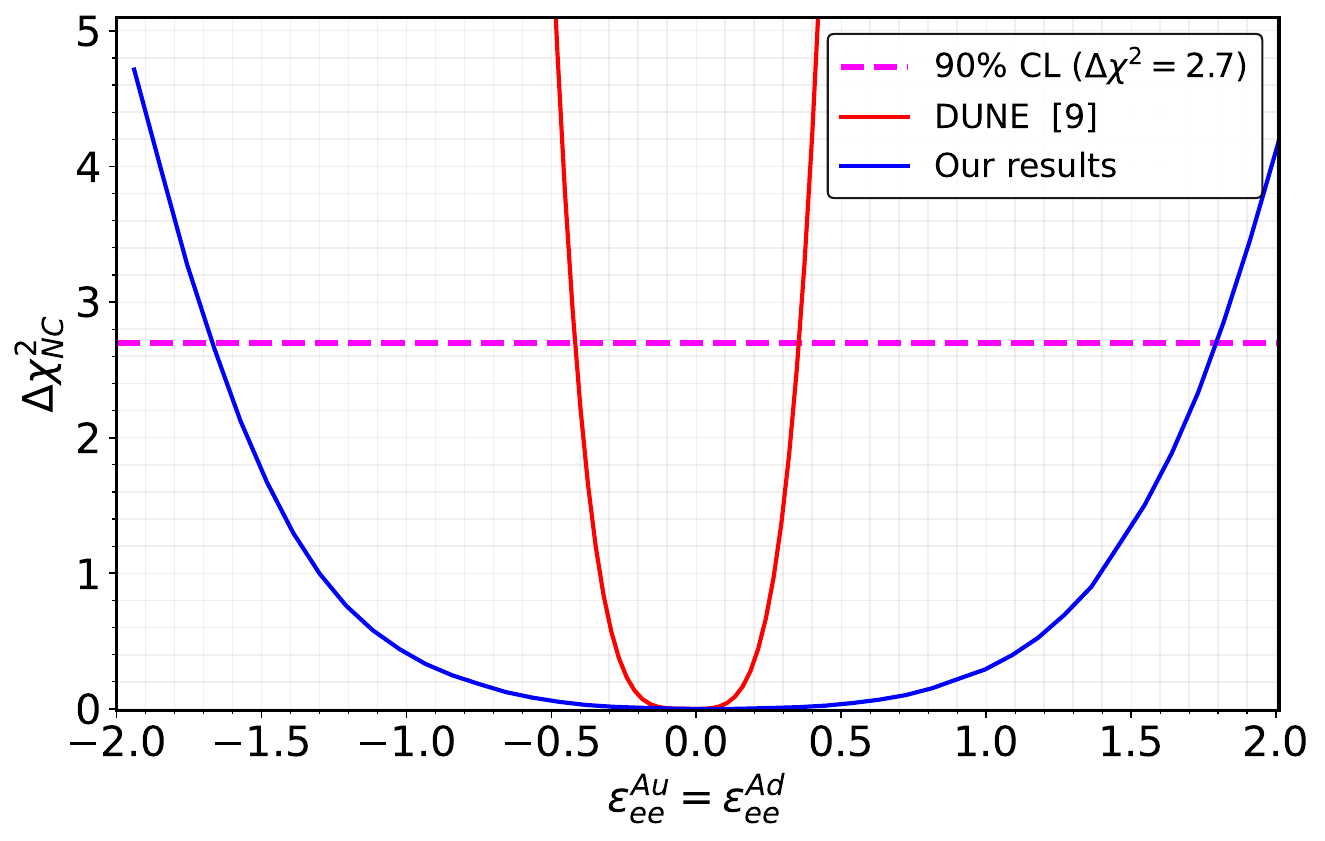}\label{fig:delta-chi2-ee-Au=Ad}}
    \subfigure[]{\includegraphics[width=0.42\textwidth ]{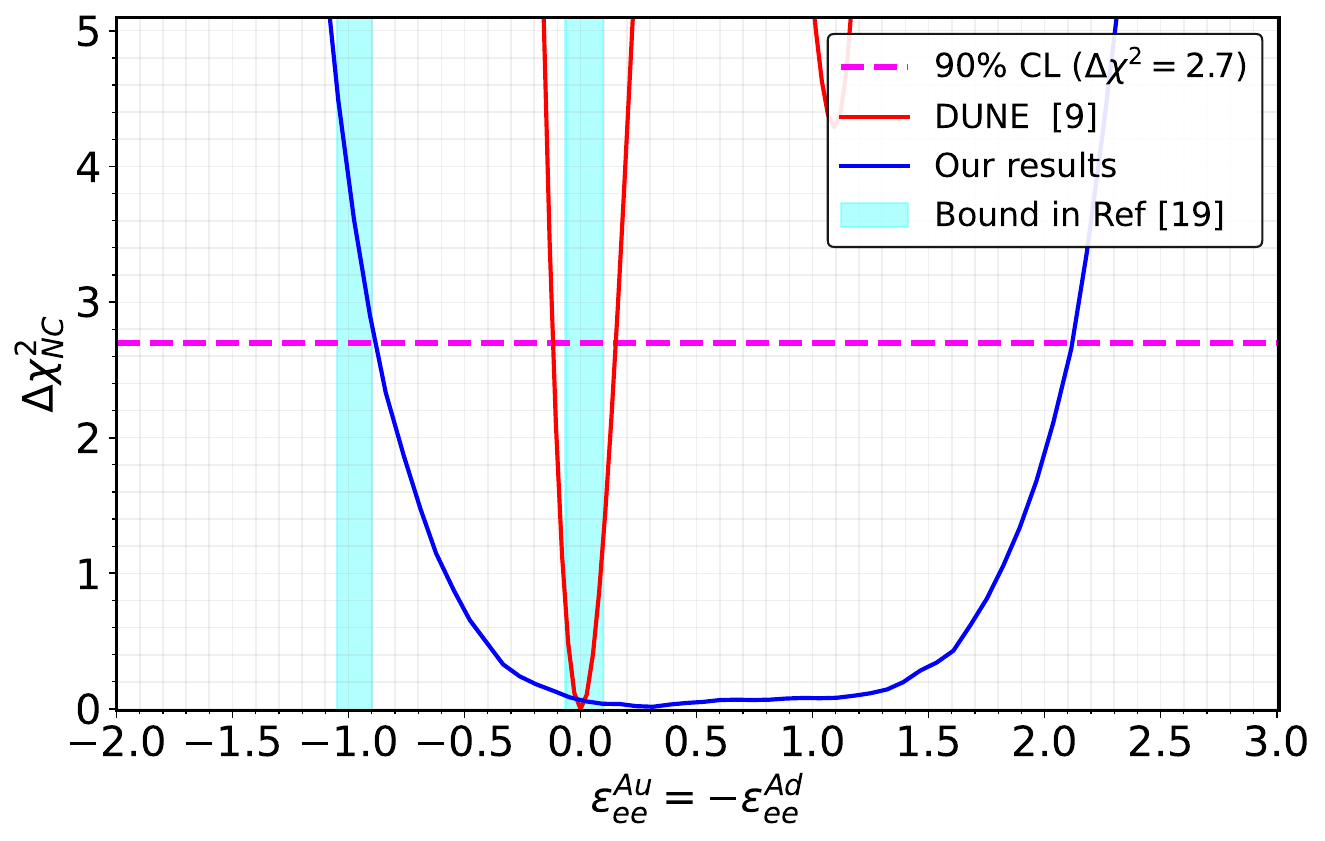}\label{fig:delta-chi2-ee-Au=-Ad}}
  \vspace{0.em}
    \caption{ $\Delta\chi^2_{NC}$ distributions for different $\epsilon^{Au}_{ee}/\epsilon^{Ad}_{ee}$ ratios. See the caption of  Fig. \ref{fig:delta-chi2-tt} for further explanation.} 
    \label{fig:delta-chi2-ee}
\end{figure*}

Figures~\ (\ref{fig:delta-chi2-tt}-\ref{fig:delta-chi2-ee}) show the $\Delta \chi_{NC}^2$ versus the NSI couplings.
The blue curves in Figs.~(\ref{fig:delta-chi2-tt}-\ref{fig:delta-chi2-ee}) show our results based on the MINOS and MINOS+ NC data bins, taking only one NSI coupling nonzero and marginalizing over the neutrino mass and mixing parameters.  $\Delta \chi^2_{NC}$ is defined in Eqs. (\ref{ChiNC},\ref{Delta-chi}). The horizontal dashed {magenta} line in these figures show $\Delta \chi^2_{NC}=2.7$, corresponding to 90 \% C.L. with one degrees of freedom.  Comparing the marginalized bounds to the bounds with  the mixing parameter fixed to the central values, we find that  the uncertainties induced by the mixing parameters  are negligible for the $\tau \tau$ and $e\tau$ components  and are relevant only for the $ee$ components. The corresponding bounds are summarized in Tab \ref{tab:bounds}.
The red curves in Figs. (\ref{fig:delta-chi2-tt}-\ref{fig:delta-chi2-ee}) show the $\Delta \chi^2_{NC}$ forecast for DUNE as found in  \cite{Abbaslu:2023vqk}.  We have shown the ideal case where the  systematic errors of DUNE are negligible. In \cite{Abbaslu:2023vqk},  we have shown how the performance of DUNE would deteriorate when   the systematic errors are turned on. Not surprisingly,  DUNE can improve on the bounds that we derive from MINOS and MINOS+. See Tab \ref{tab:bounds}, for better comparison.
In the (a), (b) and (d) panels of Figs. (\ref{fig:delta-chi2-tt}-\ref{fig:delta-chi2-ee}), we have also superimposed the 90 \% C.L. solutions presented in Ref. \cite{Coloma:2023ixt}
which are dominated by the SNO bounds. In the isosymmetric case with $\epsilon^{Au}=\epsilon^{Ad}$, the SNO bounds do not apply which means the whole (c)-panels of these figures are a solution.

As seen in  Figs (\ref{fig:delta-chi2-tt}, \ref{fig:delta-chi2-et})
and  in Tab \ref{tab:bounds}, our MINOS(+) bounds on $\epsilon_{e\tau}^{Aq}$ and $\epsilon_{\tau \tau}^{Aq}$  are stronger than the previous bounds found in \cite{Coloma:2023ixt} from SNO. Moreover, MINOS(+) rules out the non-trivial disconnected solutions found in \cite{Coloma:2023ixt} with high 
confidence level. However, for the $ee$ components, except for the case of $\epsilon^{Au}=\epsilon^{Ad}$ shown in Fig \ref{fig:delta-chi2-ee-Au=Ad}, the bounds from SNO is more stringent than the bounds that we find. This is understandable because while the $\nu_e$ contribution to the solar neutrino beam is significant, even at the far detector, the $\nu_e$ component of the MINOS(+) beam  is suppressed.

As seen from the figures, for $\epsilon^{Au} \ne \epsilon^{Ad}$ cases, there  are disconnected solutions explaining the MINOS(+) data with a $\Delta \chi^2_{NC}$ value (almost) equal to that for SM with $\epsilon^{Aq}=0$. Studying the formulas for quasi-elastic, resonance and DIS cross sections in the presence of NSI, we observe such a degeneracy. For example, from the formulas in Ref \cite{Abbaslu:2024jzo}, we observe that the cross section of the $\Delta$ resonance scattering at $\epsilon^{Au}=\epsilon^{Ad}=0$ is equal to that  at $\epsilon^{Au}=0$ and $\epsilon^{Ad}=-2$ as well as to that
 at $\epsilon^{Au}=2$ and $\epsilon^{Ad}=0$. On the other hand, from the formulas in Ref \cite{Abbaslu:2023vqk}, we find that the DIS cross section at $\epsilon^{Au}=\epsilon^{Ad}=0$ is almost equal to that at $\epsilon^{Au}=0$ and $\epsilon^{Ad}=-1$ as well as to that
 at $\epsilon^{Au}=1$ and $\epsilon^{Ad}=0$. As explained before, the energy spectrum of the MINOS(+) beam covers all these regimes. Consequently, as shown in the figures, the second minimum lies somewhere in between. According to figures \ref{fig:delta-chi2-tt}-\ref{fig:delta-chi2-ee}, SNO completely rules out the second (disconnected) solutions found by MINOS and MINOS+.

\begin{figure*}[t]
    \centering
       \subfigure[]{ \includegraphics[width=0.41\textwidth ]{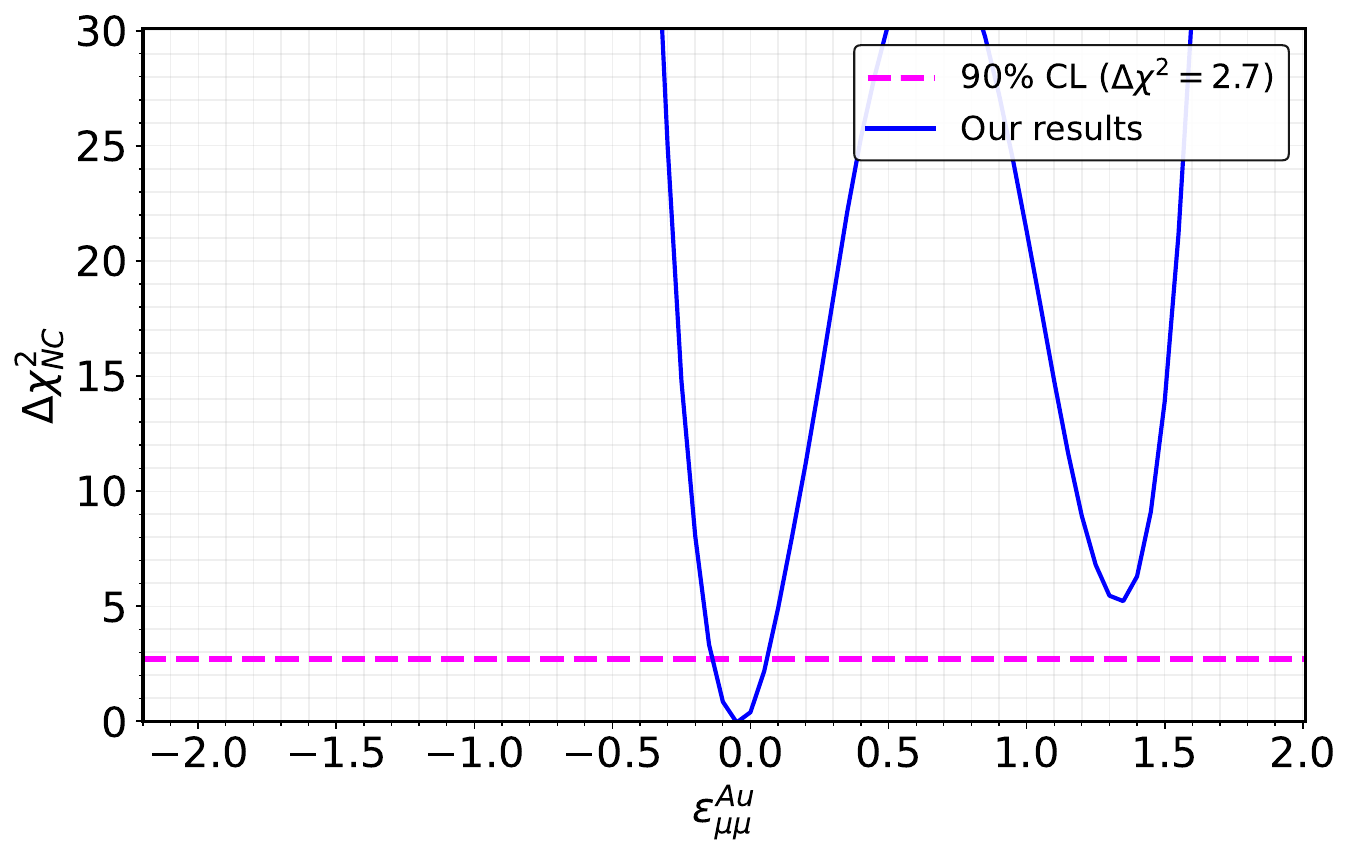}\label{fig:delta-chi2-mm-Au}}
    \subfigure[]{\includegraphics[width=0.41\textwidth ]{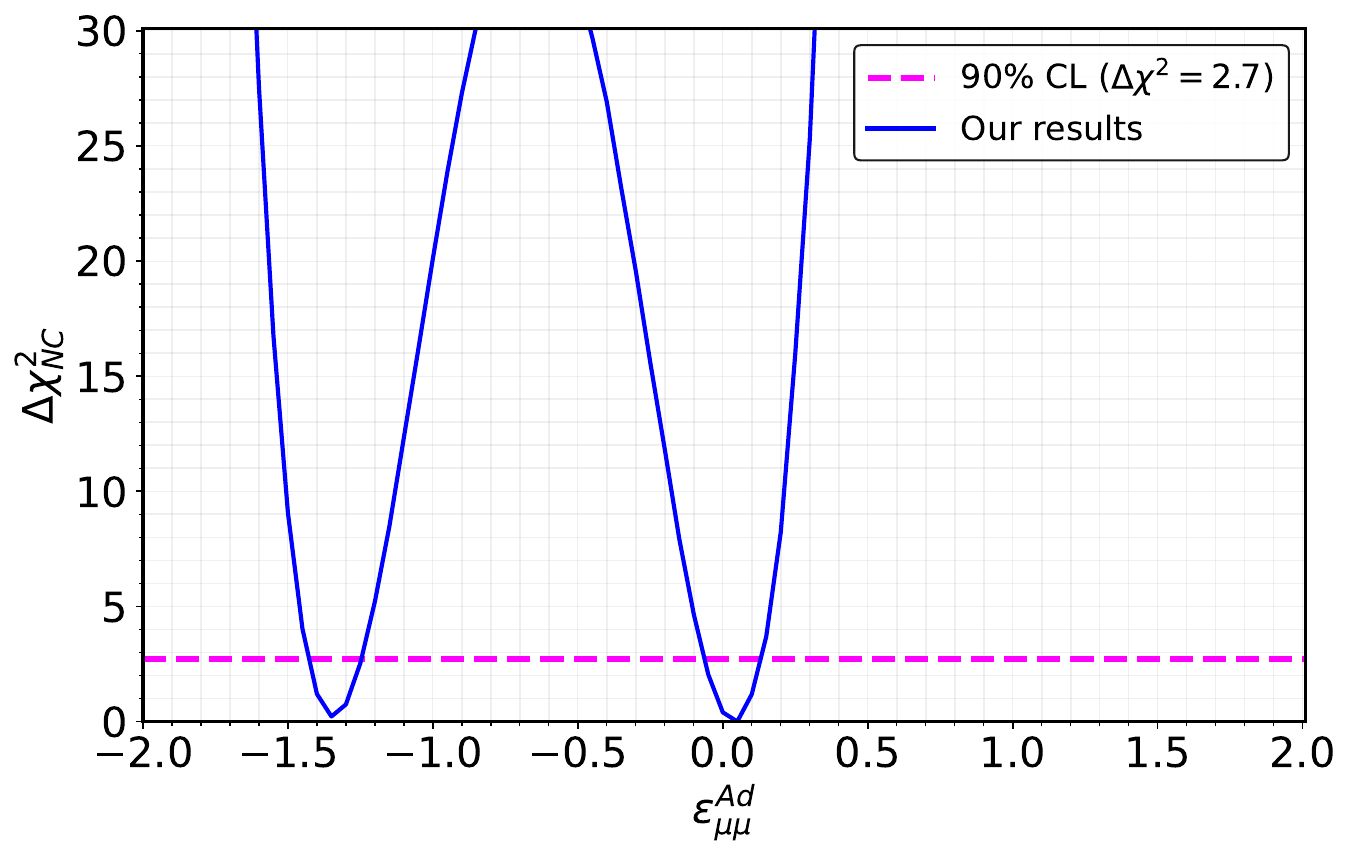}\label{fig:delta-chi2-mm-Ad}}
    \subfigure[]{\includegraphics[width=0.41\textwidth ]{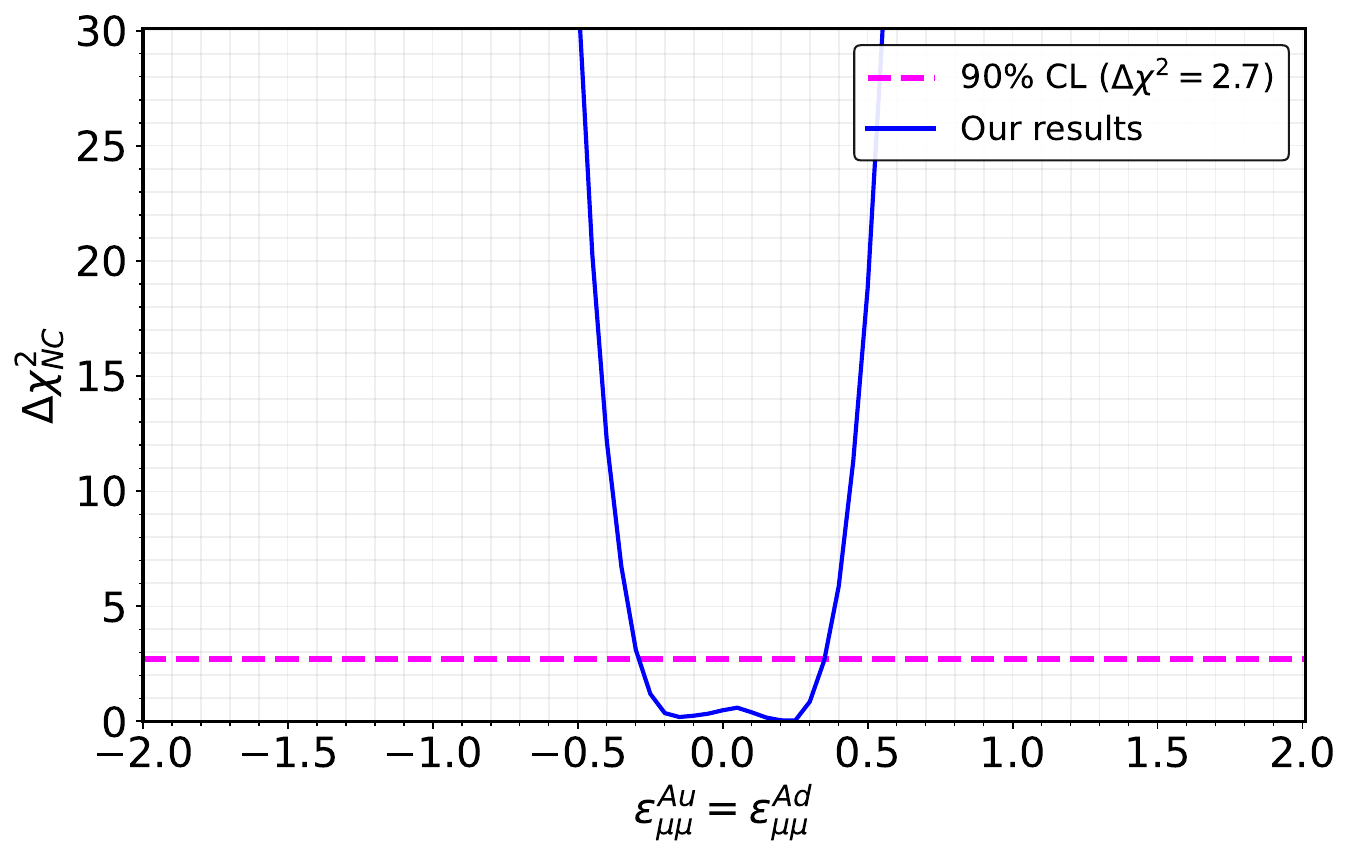}\label{fig:delta-chi2-mm-Au=Ad}}
    \subfigure[]{\includegraphics[width=0.41\textwidth ]{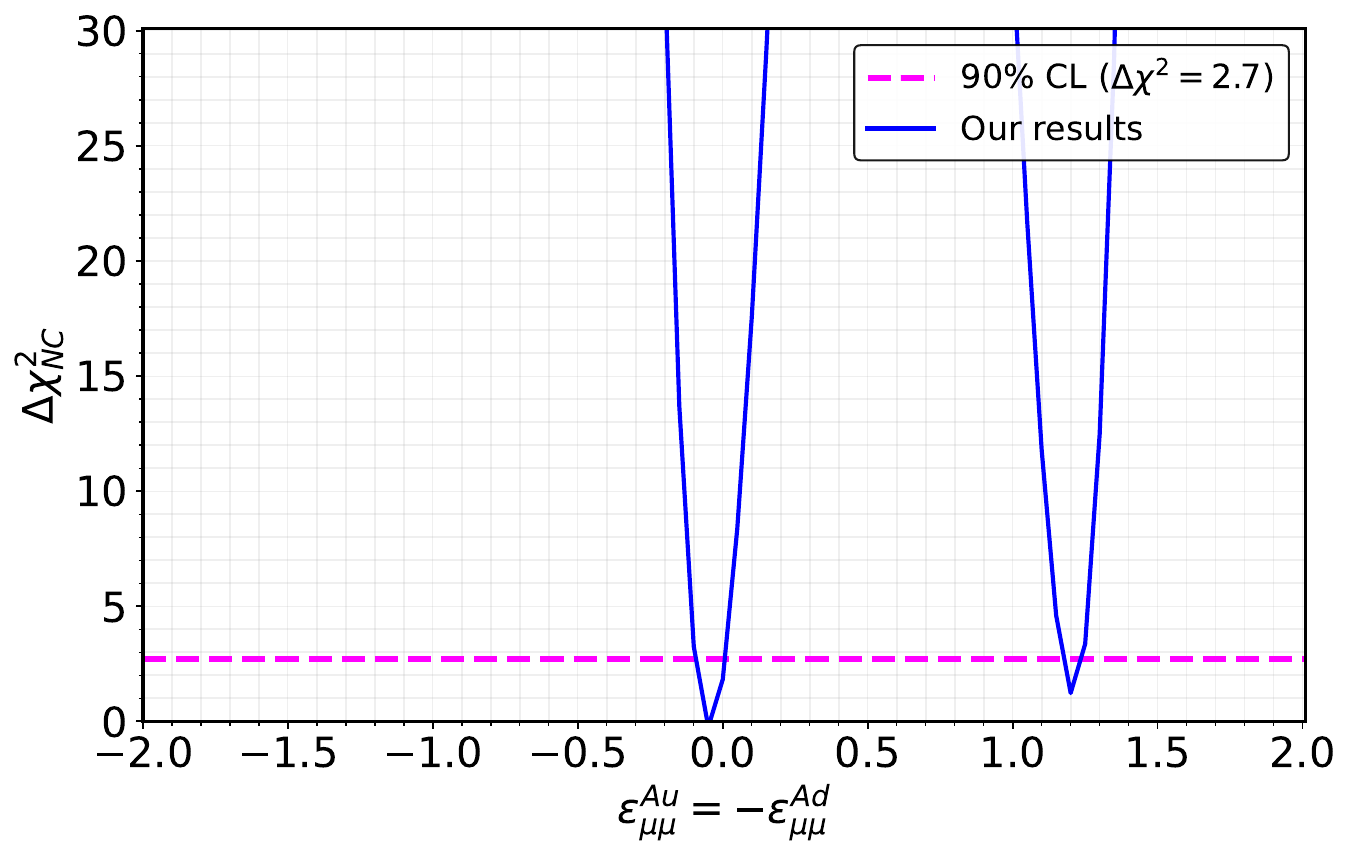}\label{fig:delta-chi2-mm-Au=-Ad}}
    \vspace{0.em}
    \caption{ $\Delta\chi^2_{NC}$ distributions for different ratios of $\epsilon^{Au}_{\mu\mu}/\epsilon^{Ad}_{\mu\mu}$. 
    The label of the horizontal axis in each panel shows the nonzero NSI
    coupling. The rest of the NSI couplings are set to zero.
   The horizontal magenta dashed lines show $\Delta \chi^2_{NC}=2.7$, corresponding to 90 \% C.L. with one degrees of freedom.  The blue curves show our results based on the MINOS and MINOS+ data.  We have marginalized over the relevant oscillation parameters using a 
Gaussian priors with widths given by the $1\sigma$ confidence intervals   as shown in Table \ref{tab:mixing-parameters}. }
    \label{fig:delta-chi2-mm}
\end{figure*}

\begin{figure*}[]
    \centering
    \subfigure[]{\includegraphics[width=0.42\textwidth ]{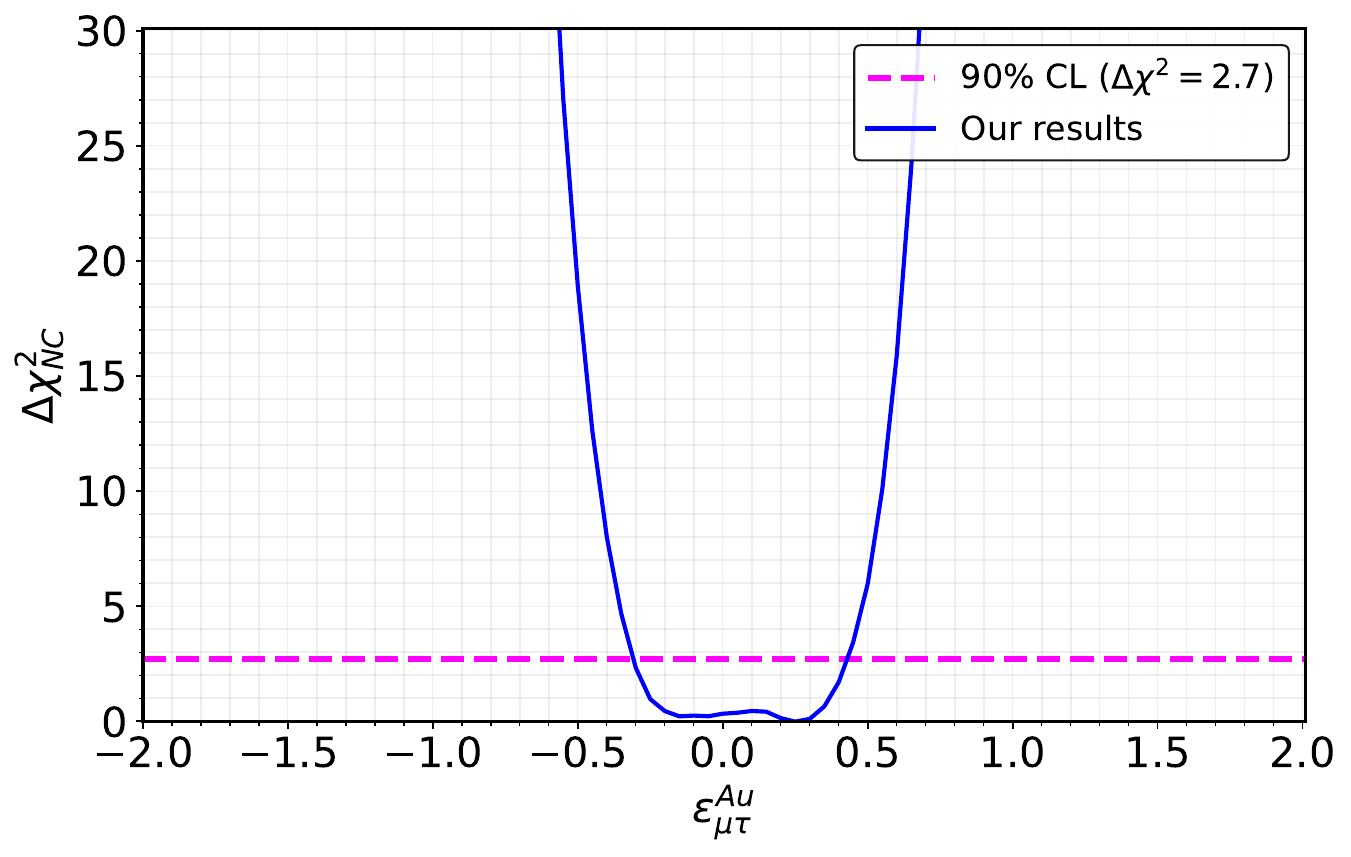}\label{fig:delta-chi2-mt-Au}}
    \subfigure[]{\includegraphics[width=0.42\textwidth ]{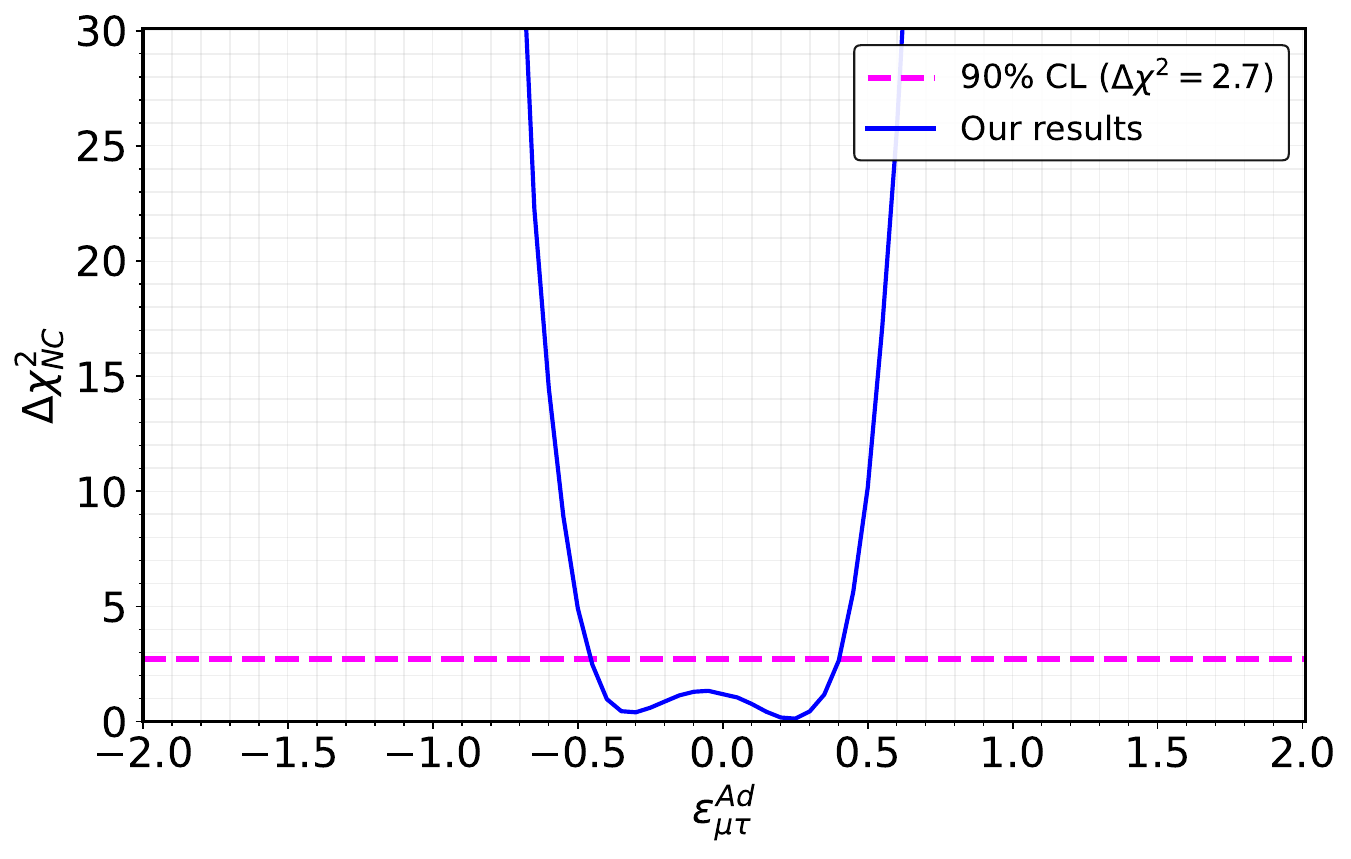}\label{fig:delta-chi2-mt-Ad}}
    \subfigure[]{\includegraphics[width=0.42\textwidth ]{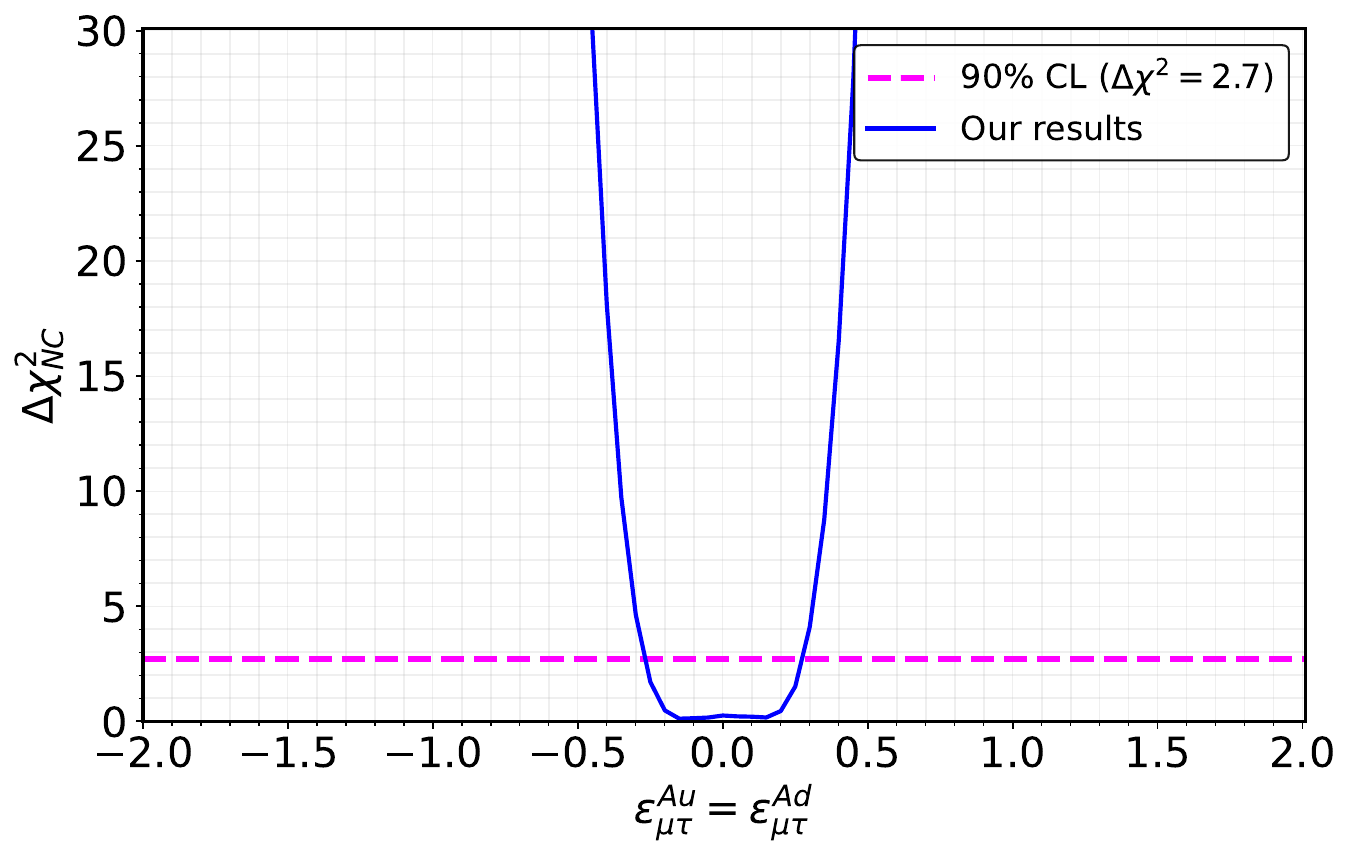}\label{fig:delta-chi2-mt-Au=Ad}}
    \subfigure[]{\includegraphics[width=0.42\textwidth ]{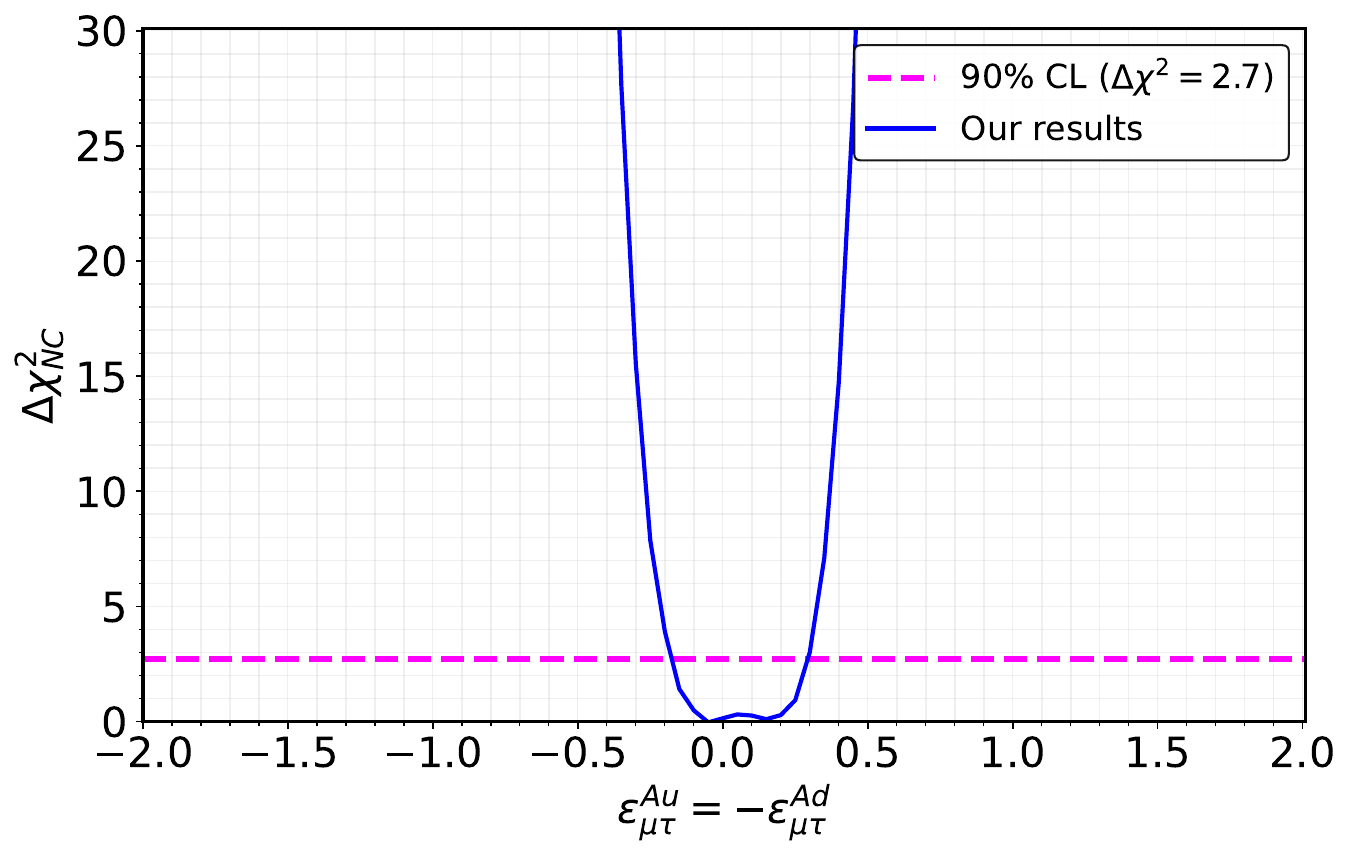}\label{fig:delta-chi2-mt-Au=-Ad}}
    \caption{ $\Delta\chi^2_{NC}$ distributions for different $\epsilon^{Au}_{\mu\tau}/\epsilon^{Ad}_{\mu\tau}$ ratios. For further explanation, see the caption of Fig.~\ref{fig:delta-chi2-mm}.}
    \label{fig:delta-chi2-mt}
\end{figure*}

Figs (\ref{fig:delta-chi2-mm},\ref{fig:delta-chi2-mt}) show our results for the $\mu \mu$ and $\mu \tau$ components, again marginalizing over the neutrino mass and mixing parameters. Even if we  fixed the neutrino mass and  mixing parameters to their central values, our results would  not noticeably change. We have not studied the $\mu e $ component as from the $\mu {\rm Ti} \to e~ {\rm Ti}$ bounds, it is severely constrained. These bounds from MINOS and MINOS+ are not competitive with the  NuTeV bounds but as discussed in \cite{Abbaslu:2023vqk}, the latter should be  taken with a grain of salt.  As seen in Tab. \ref{tab:bounds-mm-mt}, the potential reach of DUNE for the $\mu \mu$ and $\mu \tau$ components is going to be far beyond MINOS(+).

\begin{figure*}[t!]
    \centering
    {\includegraphics[width=0.98\textwidth ]{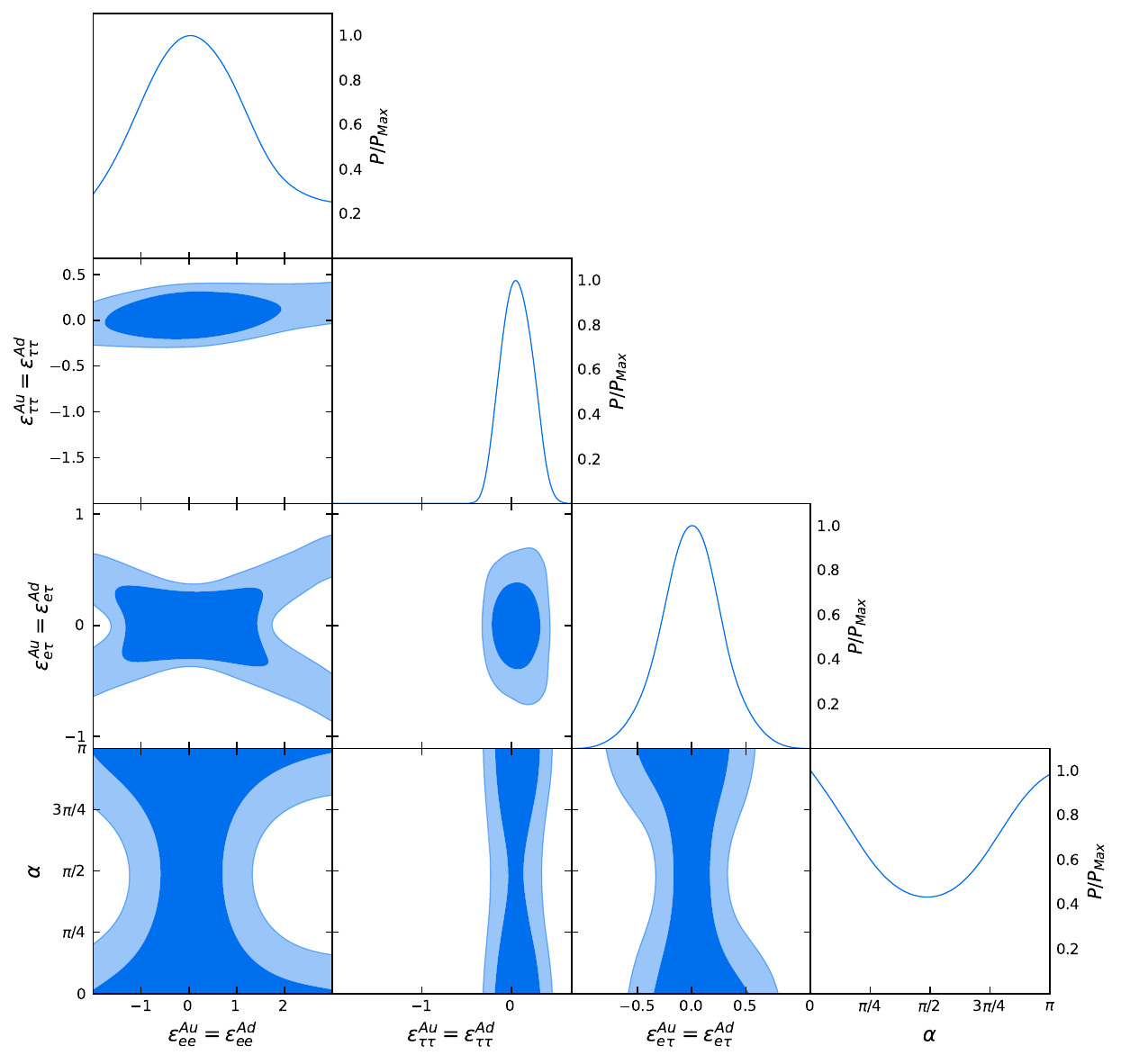} }
    \vspace{-0.8em}
    \caption{Constraints on the axial NSI parameters from the MINOS and MINOS+ data. The diagonal elements $\epsilon_{ee}$, $\epsilon_{\tau\tau}$ and the off-diagonal parameter $\epsilon_{e\tau}$ are shown, for the case of $\epsilon^{Au}_{\alpha\beta} = \epsilon^{Ad}_{\alpha\beta}$. The parameter $\epsilon_{\mu\mu}^{Au}=\epsilon_{\mu\mu}^{Ad}$ is sampled with a Gaussian prior $\mathcal{N}(0, 0.01)$. Each panel shows a two-dimensional projection of the posterior distribution after marginalization over the remaining parameters ($\Delta m^2_{32}$, $\theta_{23}$, $\theta_{13}$, $\delta$, $\epsilon_{\mu\mu}$, $\alpha$). The contours correspond to the $1\sigma$ and $2\sigma$ credible regions (2 d.o.f.).  The oscillation parameters $\theta_{12} $ and $\Delta m^2_{21}$ are fixed to their best-fit values given in Table \ref{tab:mixing-parameters}.}\label{Triangle-Au=Ad}
\end{figure*}

\begin{table*}[t]
\renewcommand{\arraystretch}{0.7}
\begin{tabular*}{\textwidth}{@{\extracolsep{\fill}}lcccl@{}}
	Coupling&Source& 90\% C.L bound \\
	\hline
	\multirow{3}{*}{$\epsilon_{ee}^{Au}$} & Ref~\ \cite{Coloma:2023ixt}& $[-2.1 , -1.8] + [-0.19 , 0.13]$  \\ 
    & DUNE \cite{Abbaslu:2023vqk}& [- 0.23 , 1.28] \\ 
    &Our results & [-1.5 , 3.02] \\
	\hline
	\multirow{3}{*}{$\epsilon_{e\tau}^{Au}$}& Ref~\ \cite{Coloma:2023ixt} & $[-1.5 , -1.3] + [-0.13 , 0.10] + [1.4 , 1.7]$ \\ 
    & DUNE  \cite{Abbaslu:2023vqk}& [-0.038 , 0.065] \\ 
	&Our results & [-0.86 , 0.22] \\
	\hline
	\multirow{3}{*}{$\epsilon_{\tau\tau}^{Au}$}  & Ref~\ \cite{Coloma:2023ixt}& $[-2.1 , -1.8] + [-0.20 , 0.15]$  \\ 
    & DUNE  \cite{Abbaslu:2023vqk}& [-0.014 , 0.014 ] \\ 
	&Our results & [-0.12 , 0.11]+[1.47 , 1.58] \\
	\hline\hline
	\multirow{3}{*}{$\epsilon_{ee}^{Ad}$}  &Ref~\ \cite{Coloma:2023ixt}& $[-0.13 , 0.19]+[1.8 , 2.1]$  \\ 
    & DUNE  \cite{Abbaslu:2023vqk} & [-1.24 , -0.93 ]$+$ [-0.35 , 0.20 ]\\ 
	&Our results & [-2.94 , 1.45] \\
	\hline
	\multirow{3}{*}{$\epsilon_{e\tau}^{Ad}$}  & Ref~\ \cite{Coloma:2023ixt}& $[-1.7,-1.4]+[-0.10,0.13]+[1.3,1.5]$  \\ 
    & DUNE  \cite{Abbaslu:2023vqk}& [-0.040 , 0.051] \\ 
	&Our results & [-0.26 , 0.82] \\ 
	\hline
	\multirow{3}{*}{$\epsilon_{\tau\tau}^{Ad}$}  & Ref~\ \cite{Coloma:2023ixt}& $[-0.15 , 0.20]+[1.8 , 2.1]$  \\ 
    &DUNE  \cite{Abbaslu:2023vqk}& [-0.016 , 0.014 ] \\ 
	&Our results & [-1.55 , -1.3] $+$[-0.12 , 0.13]\\
	\hline\hline
	\multirow{3}{*}{$\epsilon_{ee}^{Au}$=$\epsilon_{ee}^{Ad}$} & Ref~\ \cite{Coloma:2023ixt}& - \\ 
    & DUNE  \cite{Abbaslu:2023vqk}& [-0.41 , 0.35 ]\\ 
	&Our results & [-1.67 , 1.8] \\
	\hline
	\multirow{3}{*}{$\epsilon_{e\tau}^{Au}$=$\epsilon_{e\tau}^{Ad}$} & Ref~\ \cite{Coloma:2023ixt}& -  \\ 
    & DUNE  \cite{Abbaslu:2023vqk}& [-0.083 , 0.089 ] \\ 
	&Our results & [-0.32 , 0.27] \\
	\hline
	\multirow{3}{*}{$\epsilon_{\tau\tau}^{Au}$=$\epsilon_{\tau\tau}^{Ad}$} & Ref~\ \cite{Coloma:2023ixt}&-  \\ 
    & DUNE  \cite{Abbaslu:2023vqk}& [-0.084 , 0.076 ] \\ 
	&Our results & [-0.25 , 0.36] \\ 
    \hline\hline
    \multirow{3}{*}{$\epsilon_{ee}^{Au}$=-$\epsilon_{ee}^{Ad}$} & Ref~\ \cite{Coloma:2023ixt}& $ [-1.05,-0.9]+[-0.095, 0.065]$ \\ 
    & DUNE  \cite{Abbaslu:2023vqk}& [-0.12 , 0.15]\\ 
	&Our results & [-0.88 , 2.12] \\
	\hline
	\multirow{3}{*}{$\epsilon_{e\tau}^{Au}$=-$\epsilon_{e\tau}^{Ad}$} & Ref~\ \cite{Coloma:2023ixt}&$[-0.75, -0.65]+[-0.065, 0.05]+[+0.7, 0.85]$  \\ 
    & DUNE  \cite{Abbaslu:2023vqk}& [-0.023 , 0.017] \\ 
	&Our results & [-0.65 , -0.37]$+$[-0.17 , 0.12] \\
	\hline
	\multirow{3}{*}{$\epsilon_{\tau\tau}^{Au}$=-$\epsilon_{\tau\tau}^{Ad}$} & Ref~\ \cite{Coloma:2023ixt}& $[-1.05,-0.9]+[-0.095,+0.075] $  \\ 
    & DUNE  \cite{Abbaslu:2023vqk}& [-0.003 , 0.003] \\ 
	&Our results & [-0.05 , 0.05]$+$[1.18 , 1.28]\\ 
    \hline
\end{tabular*}
\caption{The 90\% C.L. bounds on the $ee$, $e\tau$, and $\tau\tau$ components  for various values of the ratio $\epsilon_{\alpha\beta}^{Au}/\epsilon_{\alpha\beta}^{Ad}$. Our results are based on the MINOS+ and MINOS NC data,  taking into account the  uncertainties  of the neutrino parameters shown in Tab. \ref{tab:mixing-parameters}. The shown 90 \% C.L. ranges  for DUNE  do not include  systematic errors   \cite{Abbaslu:2023vqk}. The ranges from Ref. \cite{Coloma:2023ixt} are mainly based on the SNO NC data.}\label{tab:bounds}
\end{table*}

\begin{table*}[t]
\renewcommand{\arraystretch}{0.7}
\begin{tabular*}{\textwidth}{@{\extracolsep{\fill}}lcccl@{}}
\hline
	Coupling&Source& 90\% C.L bound \\
	\hline
	\multirow{2}{*}{$\epsilon_{\mu\mu}^{Au}$} 
    & DUNE \cite{Abbaslu:2023vqk}& $[-4\times10^{-7},4\times10^{-7}]$\\ 
    &Our results &$[-0.14 , 0.06]$\\
	\hline
	\multirow{2}{*}{$\epsilon_{\mu\tau}^{Au}$}
    & DUNE \cite{Abbaslu:2023vqk}&$[-5\times10^{-4},5\times10^{-4}]$\\ 
	&Our results &$[-0.3 , 0.42]$\\

	\hline\hline
	\multirow{2}{*}{$\epsilon_{\mu\mu}^{Ad}$}  
    & DUNE \cite{Abbaslu:2023vqk}& $[ -3\times10^{-7},3\times10^{-7}]$\\ 
	&Our results &$[-1.4 , -1.2]+[-0.06 , 0.13]$\\
	\hline
	\multirow{2}{*}{$\epsilon_{\mu\tau}^{Ad}$} 
    & DUNE \cite{Abbaslu:2023vqk}&$[-4.5\times10^{-4},4.5\times10^{-4}]$\\ 
	&Our results &$[-0.42 , 0.33]$\\ 
	
	\hline\hline
	\multirow{2}{*}{$\epsilon_{\mu\mu}^{Au}$=$\epsilon_{\mu\mu}^{Ad}$}
    & DUNE \cite{Abbaslu:2023vqk}& $[-2\times10^{-5},2\times10^{-5}]$\\ 
	&Our results &$[-0.29 , 0.35]$\\
	\hline
	\multirow{2}{*}{$\epsilon_{\mu\tau}^{Au}$=$\epsilon_{\mu\tau}^{Ad}$}
    & DUNE \cite{Abbaslu:2023vqk}& $[-1\times10^{-4},1\times10^{-4}] $\\ 
	&Our results &$[-0.18 , 0.18]$\\
    \hline\hline
    \multirow{2}{*}{$\epsilon_{\mu\mu}^{Au}$=-$\epsilon_{\mu\mu}^{Ad}$} 
    & DUNE \cite{Abbaslu:2023vqk}& $[-1\times10^{-7},1\times10^{-7}]$\\ 
	&Our results &$[-0.09 , 0.007]+[1.18 , 1.24]$\\
	\hline
	\multirow{2}{*}{$\epsilon_{\mu\tau}^{Au}$=-$\epsilon_{\mu\tau}^{Ad}$} 
    & DUNE \cite{Abbaslu:2023vqk}&$[-1\times10^{-4},1\times10^{-4}]$\\ 
	&Our results &$[-0.17 , 0.29]$\\ 
    \hline
\end{tabular*}
\caption{The 90\% C.L. bounds on the $\mu\mu$ and $\mu\tau$ components for various values of the ratio $\epsilon_{\alpha\beta}^{Au}/\epsilon_{\alpha\beta}^{Ad}$. Our results are based on the MINOS+ and MINOS NC data,  taking into account the  uncertainties  of the neutrino parameters shown in Tab. \ref{tab:mixing-parameters}. The shown 90 \% C.L. ranges  for DUNE  do not include  systematic errors   \cite{Abbaslu:2023vqk}. }\label{tab:bounds-mm-mt}
\end{table*}

Let us now consider the case that more than one NSI coupling is nonzero. As discussed before, there are already strong bounds on the isovector combination, $\epsilon^{Au}=-\epsilon^{Ad}$. As seen from Tables (\ref{tab:bounds},\ref{tab:bounds-mm-mt}), even for single nonzero NSI with $\epsilon^{Au}=-\epsilon^{Ad}$, the improvement of the bound by MINOS(+)  is not significant. When we allow for more than one element of $\epsilon^{Au}_{\alpha \beta}=-\epsilon^{Ad}_{\alpha \beta}$ to be nonzero, the bounds from MINOS(+) become irrelevant in comparison to the already existent bounds.  We have therefore focused on the isosinglet case, $\epsilon^{Au}=\epsilon^{Ad}$ and have shown our results in Fig. \ref{Triangle-Au=Ad}. We apply the strong bounds on $\epsilon_{\mu \mu}^{Au}=\epsilon^{Ad}_{\mu \mu}$. Similarly, we apply Gaussian priors for the relevant mixing parameters, $\theta_{23}$, $\Delta m_{31}^2$, $\theta_{13}$ and $\delta$ with the central values and width given in Table \ref{tab:mixing-parameters}. In our analysis, the phase of $\epsilon_{\tau e}$, $\alpha$ is allowed to vary between 0 and $\pi$ with $\epsilon_{e \tau}$ and  $\epsilon_{\tau \tau}$ taking both positive and negative values to cover the whole physical range. Comparing the results shown in Table \ref{tab:bounds} and in Fig. \ref{Triangle-Au=Ad}, we observe that the  bound on $\epsilon^{Au}_{\tau\tau}=\epsilon^{Ad}_{\tau \tau}$ only slightly weakens when we allow all the components to be simultaneously nonzero. However, the panel showing $\epsilon_{e\tau}$ versus $\epsilon_{ee}$ demonstrates a peculiar behavior with extended legs at 2$\sigma$. These extended legs can be explained as follows. Consider a $|\nu_{far}\rangle$ state with a definite energy and therefore given $\mathcal{A}_e$, $\mathcal{A}_\mu$ and $\mathcal{A}_\tau$. If the flavor structure of the NSI is perpendicular to $|\nu_{far}\rangle$ (which requires $\epsilon_{ee} \mathcal{A}_e+\epsilon_{e\tau} e^{i \alpha}\mathcal{A}_\tau=0 $), the neutrino will not feel NSI. The legs correspond to $\epsilon_{ee}/\epsilon_{e \tau}=\pm |\mathcal{A}_\tau|/|\mathcal{A}_e|$ with $\pm$ coinciding with $\alpha=0,\pi$. This also explains the sand clock shape of the $\alpha$ versus $\epsilon_{ee}$ panel. Considering that $\mathcal{A}_\alpha$ are functions of energy but $\epsilon_{\alpha \beta}$ are constant, this condition cannot hold for all energies so the degeneracy is not complete and the legs appear only at 2$\sigma$. As seen from these figures, $\alpha$ is unconstrained by MINOS(+).

 We also studied the possible bounds on $\epsilon^{Vs}$ and $\epsilon^{As}$. Not surprisingly,  the MINOS and MINOS+ constraints on the $ee$, $e \tau$ and $\tau \tau$ components are not  better than $O(1)$. Because of the large $\nu_\mu$ statistics in the near detector, we naively expect strong bounds on $\epsilon^{Vs}_{\mu\mu}$ and $\epsilon^{As}_{\mu \mu}$ but we find that the bounds from MINOS(+) do not turn out to be better than $O(1)$ in this case, either. The reason is that the deviations caused by $\epsilon^{Vs}_{\mu \mu}$ and $\epsilon^{As}_{\mu \mu}$ in the near detector bins can also be mimicked by uncertainties such as that in the flux normalization. If the normalization of the flux was fixed by the CC interactions, improvements on the  $\epsilon^{Vq}_{\mu\mu}$ and $\epsilon^{Aq}_{\mu\mu}$ bounds would be possible but the collaboration has not provided the covariance matrix between the NC and CC bins. Such information in the future experiments would be tremendously useful for searching for new physics.

\section{Summary and discussion \label{sec:Sum}}
We have studied the NC events collected by MINOS and MINOS+ far and near detectors to probe axial Non-Standard Interactions (NSI) of neutrinos with the $u$ and $d$ quarks. Since the energy spectra of MINOS and MINOS+ cover all scattering regimes from the quasi-elastic to resonance and then to DIS, we have included all these regimes in our computation. We have used NuWro \cite{nuwro_github,Juszczak:2005zs,Golan:2012rfa,Golan:2012wx} to account for the nuclear effects in the scattering.  Tab. \ref{tab:bounds} shows the 90 \% C.L. bounds on the $ee$, $e\tau$ and $\tau\tau$ components of the NSI couplings, comparing them with the previous bounds in the literature and the future reach of an ideal  DUNE-like experiment with negligible systematic errors. 
Before the present paper, the $ee$, $\tau\tau$ and $\tau e$ components of the isospin singlet NSI ({\it i.e.,} $\epsilon_{\alpha \beta}^{Au}=\epsilon_{\alpha \beta}^{Ad}$ with $\alpha,\beta \in \{ e, \tau\}$) were unconstrained. We have found that MINOS and MINOS+ constrain the absolute values of the $e\tau$ and $\tau \tau$ components of isospin singlet NSI to be smaller than $\sim 0.3$.  We have also studied the degeneracies among the NSI couplings when they are simultaneously allowed to be nonzero. The results are shown in Fig.~\ref{Triangle-Au=Ad}.

For the general values of  $\epsilon^{Au}/\epsilon^{Ad}$, the bound from MINOS and MINOS+ on the $\tau\tau$ components  surpass the previous bounds in the literature. The MINOS(+) bounds on the $ee$ and $e\tau$ components are not competitive  with the previous bounds for $\epsilon^{Au}/\epsilon^{Ad}\ne 1$ but, with a high confidence level, MINOS(+) rules out the non-trivial disconnected solutions that were found in the previous data. 
We have also derived the bounds on $\epsilon_{\mu \alpha}^{Aq}$ from the MINOS(+) data but they are not competitive with the present bounds 
\cite{Davidson:2003ha}. Moreover, we studied the effects of both vector and axial non-standard interaction of neutrinos with the $s$ quark. The bounds from 
MINOS(+) on $\epsilon^{Vs}$ and $\epsilon^{As}$ are of order of $O(1)$.

As discussed in \cite{Abbaslu:2023vqk}, 
   the future DUNE experiment can probe much smaller values of the axial NSI. If an excess (rather than a deficit) of
the NC events at DUNE relative to the SM prediction is discovered, it would be a strong hint in favor of NSI within a certain range of couplings because most of alternative beyond standard model scenarios such as the oscillation to the sterile neutrinos predict only a deficit of the NC events.

In \cite{Abbaslu:2024jzo}, we had suggested to use the low energy atmospheric neutrino data collected by KamLAND to probe $\epsilon_{\tau\tau}^{Aq}$ 
down to $\sim 0.4$. In this paper, we have found that such values are already ruled out by MINOS and MINOS+. However, the same idea can be invoked by the JUNO collaboration to probe smaller values of $\epsilon^{Aq}_{\tau \tau}$.

\section*{Acknowledgments} 
We would like to thank A. Aurisano and A. Sousa from the MINOS collaboration for the clarification about the supplementary material of \cite{MINOS:2017cae}. We would like also to thank C. Gonzalez-Garcia for useful discussions and the  collaboration in the early stages of the project. {We are very grateful to S. Ansarifard for his valuable comments, insightful discussions, and guidance on the MCMC code.} We are also grateful to H. Abdolmaleki and  J. P. Pinheiro, for useful discussions.
This work is based on research funded by Iran National Science Foundation
(INSF) under project No.4031487.
 This project has received funding from the European Union’s Horizon Europe research and innovation programme under the Marie Skłodowska-Curie Staff Exchange grant agreement No 101086085 – ASYMMETRY.
 The authors acknowledge the computational resources provided by the SARV computing facility at the school of theoretical physics of IPM.
\bibliography{biblio}

\end{document}